\documentclass[fleqn,usenatbib]{mnras}

\usepackage{newtxtext,newtxmath}

\usepackage[T1]{fontenc}

\DeclareRobustCommand{\VAN}[3]{#2}
\let\VANthebibliography\thebibliography
\def\thebibliography{\DeclareRobustCommand{\VAN}[3]{##3}\VANthebibliography}

\usepackage{graphicx}	
\usepackage{amsmath}	
\usepackage{float}
\usepackage{subcaption}
\usepackage{xcolor}
\usepackage{xspace}
\usepackage{soul,xcolor}

\newcommand{\kms}{$\mathrm{km\ s^{-1}}$\xspace}	

\title[LOSVDs in edge-on galaxies]{Uncovering the origin of complex velocity distributions in edge-on disc galaxies}

\author[S.Townson et al.]{
Sophie Townson$^{1}$\thanks{E-mail: S.L.Townson@2021.ljmu.ac.uk},
Marie Martig$^{1}$,
Robert J.J. Grand$^{1}$,
Prashin Jethwa$^{2}$,
Jes{\'u}s Falc{\'o}n-Barroso$^{3,4}$ and 
\newauthor
Timothy A. Davis$^{5}$
\\
$^{1}$Astrophysics Research Institute, Liverpool John Moores University, 146 Brownlow Hill, Liverpool L3 5RF, UK\\
$^{2}$Institute for Astronomy (IfA), University of Vienna, Türkenschanzstrasse 17, A-1180 Vienna, Austria\\
$^{3}$Instituto de Astrof{\'i}sica de Canarias, Calle Via L{\'a}ctea s/n, 38200 La Laguna, Tenerife, Spain\\
$^{4}$Depto. Astrof{\'i}sica, Universidad de La Laguna, Calle Astrof{\'i}sico Francisco S{\'a}nchez s/n, 38206 La Laguna, Tenerife, Spain\\
$^{5}$Cardiff Hub for Astrophysics Research \& Technology, School of Physics \& Astronomy, Cardiff University, Queens Buildings, Cardiff, CF24 3AA, UK\\
}

\date{Accepted 2026 October 1. Received 2026 September 23; in original form 2026 August 18}

\pubyear{\the\year{}}

\begin{document}
\setstcolor{red}
\label{firstpage}
\pagerange{\pageref{firstpage}--\pageref{lastpage}}
\maketitle

\begin{abstract}
Stellar line-of-sight velocity distributions are a powerful tool for studying galaxy kinematics and formation history, with multiple-peaked LOSVDs often indicating the presence of complex kinematic components. We extract stellar LOSVDs across the disc
of a simulated spiral galaxy viewed edge-on, and study how often and why we see multiple-peaked LOSVDs. We find that the multiple-peaked LOSVDs are mainly found close to the midplane because they result from the presence of spiral arms. This is not because the spiral arms have significantly different kinematics to the disc, but because they correspond to over-densities. Variations in the number of arms present along a line-of-sight create variations in the number of peaks in LOSVDs. This can be used to quantify the strength of the spiral arms above the midplane, and how the thickness of spiral arms vary across the disc. We find that the commonly used fourth order Gauss-Hermite moment, $h_4$, is negative in the regions where there are multiple-peaked LOSVDs, and hence traces spiral arms. Finally, we discuss the observability of the multiple-peaked regions and their associated $h_4$ measurements with current integral field spectrographs. We find that the spectral resolution of MUSE might not be high enough to uncover multiple peaks in LOSVDs when they result from spiral arms, but should allow for identification of the regions of negative $h_4$.
On the other hand, higher resolution spectrographs like KOALA and WEAVE may be able to recover multiple peaks in LOSVDs, representing a promising avenue for characterising spiral arms in edge-on disc galaxies.

\end{abstract}

\begin{keywords}
galaxies: kinematics and dynamics -- galaxies: spiral -- methods: numerical
\end{keywords}



\section{Introduction}

Line of sight velocity distributions (LOSVDs), specifically stellar LOSVDs, represent the distribution of stars as a function of their velocity along a particular line-of-sight. They are a powerful tool for uncovering information about a galaxy's kinematics and formation history. In the simple case of an edge-on axisymmetric exponential disc, the LOSVD along any given line-of-sight includes contributions from stars at a range of radii: this includes stars that are in the middle of the disc (on the line of nodes) as well as stars in the front and in the back of the galaxy. On the line of nodes, stars on circular orbits move directly along the line-of-sight; this results in a maximum line-of-sight velocity in these regions \citep{Wagner, Kregel2005}. As the distance from the line of nodes increases, the projected line-of-sight velocity of stars decreases: their orbit is not directly aligned with the line-of-sight. In addition, density is also highest along the line of nodes. This means that  the LOSVD has a peak at high absolute velocity corresponding to stars on the line of nodes, and a tail towards the galaxy systemic velocity coming from stars in front of and behind the line of nodes.

LOSVDs can also exhibit complex shapes. If stars on eccentric orbits exist, LOSVDs can broaden or contain long tails towards low or high velocities. LOSVDs may also contain multiple peaks, often interpreted as arising from kinematically diverse components within galaxies. Striking cases are galaxies such as NGC 4550 \citep{Rix1992, johnston2013disentangling} and IC 719 \citep{katkov2013}, which host counter rotating discs, producing double-peaked LOSVDs. Triple-peaked LOSVDs can also be a result of multiple kinematically distinct components coexisting, as in in the central region of FCC 47 \citep{Lamprecht2026}.

Although the extraction of multiple-peaked LOSVDs is achievable through non-parametric methods \citep{1992Rix, 1994saha, 2000Gebhardt,2021Falcon,2023Merghan,2026Prashin}, this is rarely done. In many cases, the LOSVD is instead described by a parametric function, most often using Gauss-Hermite polynomials \citep{van_der_Marel, 1993Gerhard}.  The Gauss-Hermite coefficients measure how much an LOSVD deviates from a true Gaussian distribution. The third order coefficient h$_3$ is a measure of how asymmetric the distribution is. A tail towards larger velocities results in a positive h$_3$ and a tail towards lower velocities results in a negative h$_3$. The fourth order coefficient h$_4$ is a measure of how peaked the distribution is and how broad the wings are. A more flat-topped distribution with small wings results in a negative h$_4$ and a more peaked distribution with broad wings results in a positive h$_4$. 

Such kinematic measurements are extremely useful in edge-on galaxies as a way to better separate the bulge, disc, and bar components. In an edge-on, axisymmetric rotating disc, $h_3$ is anti-correlated with velocity \citep{van_der_Marel, 1993Gerhard}. This anti-correlation is seen within the main discs of galaxies, but is also a way to identify nuclear discs \citep{Guerou,Pinna, Martig2021,Fraser_McKelvie_2025}. When a bar is present,  $h_3$ and velocity become correlated within the extent of the bar \citep{Fisher, Bureau2005, Molaeinezhad_2015}.  The strength of these signatures depends on bar strength and viewing angle \citep{Iannuzzi_2015, Zhao-lu}. There is also complexity in the $h_4$ maps for galaxies that contain a bar, with deep $h_4$ minima over the region of the bar \citep{Chung_2004,Iannuzzi_2015, Zhao-lu, Bureau2005}. 

Spiral arms are not usually discussed in the context of stellar LOSVDs, but have been detected in observations of gas in edge-on disc galaxies. \cite{Kregel2004} shows that P-V diagrams of HII regions trace the presence of spiral arms and can be used to reconstruct their position across the disc. Some other studies have used molecular gas, e.g. \cite{jim} finds evidence for a spiral structure in the P-V diagram of CO gas in the edge-on disc galaxy NGC 891, with the spiral features appearing at lower velocities. 

Ultimately, studying the kinematics of stars in edge-on disc galaxies may allow for a more precise classification of these galaxies within the Hubble sequence, whilst also allowing us to uncover the kinematic imprint of spiral structure on galactic discs. This becomes particularly relevant given the increasing availability of kinematic information from nearby edge-on disc galaxies, provided by surveys such as GECKOS  \citep{2024jesse}. In this paper, we examine the detailed shape of stellar LOSVDs throughout the disc of an edge-on simulated galaxy. We aim to study where complex LOSVDs arise and how we can use these to probe the structure of edge-on galaxies. In Section 2, we describe the simulation we use and the methods we developed to extract LOSVDs and characterise their shape. In Section 3, we explore how frequent multiple-peaked LOSVDs are and show that they can be used to trace the  presence of spiral arms. In Section 4, we discuss the observability of these signatures, and in Section 5 we present our conclusions.

\section{Simulations}
The simulation used in this paper is part of a suite of 33 zoom cosmological re-simulations first presented in \cite{Martig2012}. In this Section, we first describe the simulation technique, then present the galaxy we chose for this analysis, and finally describe the method we used to study its kinematics.

\begin{figure}
    \centering
    \includegraphics[width=\linewidth]{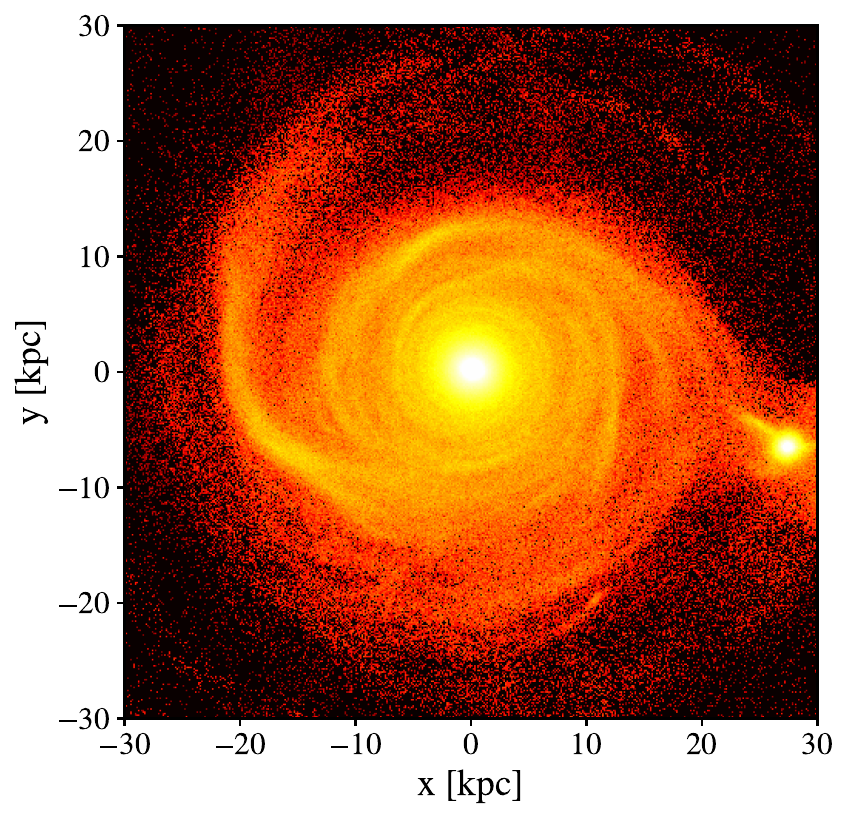}

    \caption{Mock $g$-band image of Galaxy 102 viewed face-on (see \citealp{Martig2012}). The galaxy has a strong spiral structure that is excited by the satellite seen on the right of the image.}
    \label{fig:galaxy102}
\end{figure}

\subsection{Simulation technique}
The simulations were built using a zoom-in resimulation technique consisting of two steps (see Appendix A of \citealp{Marie2009}). First, a dark matter-only cosmological simulation is run, and accretion histories are extracted for some target halos. Then, those histories are re-simulated at high resolution in new simulations containing both dark matter and baryons.

The  dark matter-only cosmological simulation is carried out with the adaptive mesh refinement code, RAMSES \citep{ADAPTIVE}, using $512^3$ dark matter particles within a box of comoving length $20 h^{-1}$ Mpc. It adopts a $\Lambda$CDM cosmology with $\Omega_m = 0.3$, $\Omega_\Lambda = 0.7$, $H_0 = 70\ \mathrm{kms^{-1}Mpc^{-1}}$ and $\sigma_8 = 0.9$. Within this simulation, we choose 33 target halos with a final mass between $2.7\times 10^{11}$ and
$2\times 10^{12}$ M$_\odot$ that live in isolated environments. We record their formation histories from $z=5$ to $z=0$: a fixed spherical boundary is placed around the halos and every halo or diffuse particle crossing that boundary is recorded. 

Those histories are then used to run 33 new individual simulations following the evolution of seed galaxies from $z=5$ to $z=0$. In these simulations, incoming dark matter halos are replaced with new galaxies containing dark matter, gas and stars, and every diffuse dark matter particle is replaced with a `blob' of higher resolution gas and dark matter particles. These simulation are run using a Particle-Mesh code where gas dynamics is modelled with a `sticky particles' algorithm \citep{ParticleMesh2002}. The spatial resolution is 150 pc. The box size is $800 \times 800 \times 800$ kpc, and the mass resolution is  $3.0\times10^5$ M$_{\odot}$ for dark matter particles, $1.5\times10^4$ M$_{\odot}$ for gas particles, $7.5\times 10^4$  M$_{\odot}$ for star particles present in seed galaxies and $1.5\times10^4$ M$_{\odot}$ for star particles formed during the simulation.

Star formation is modelled using the Schmidt–Kennicutt law \citep{kennicutt}, with an exponent of 1.5 and an efficiency of 2\%. The threshold for star formation  is set at $0.03\ \mathrm{M_{\odot}pc^{-3}}$. Energy feedback from supernovae is also included in the simulations, with $20\%$ of the energy being distributed as kinetic energy to the neighbouring gas particles, and stellar mass loss  is implemented as described in \cite{2010Martig}.

\subsection{Galaxy 102}
\begin{figure*}
    \centering
    
    \begin{subfigure}{0.49\linewidth}
        \centering
        \includegraphics[width=\linewidth]{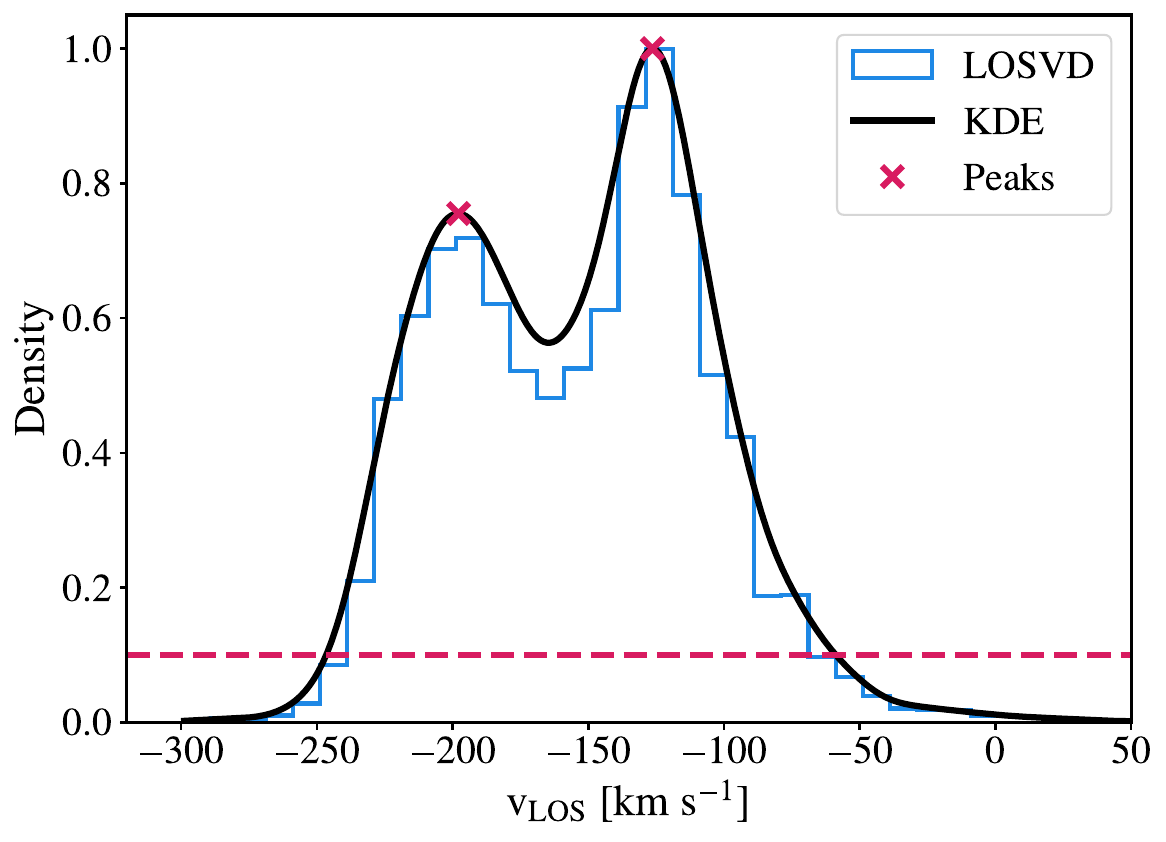}
       
    \end{subfigure}
    \hfill
    \begin{subfigure}{0.49\linewidth}
        \centering
        \includegraphics[width=\linewidth]{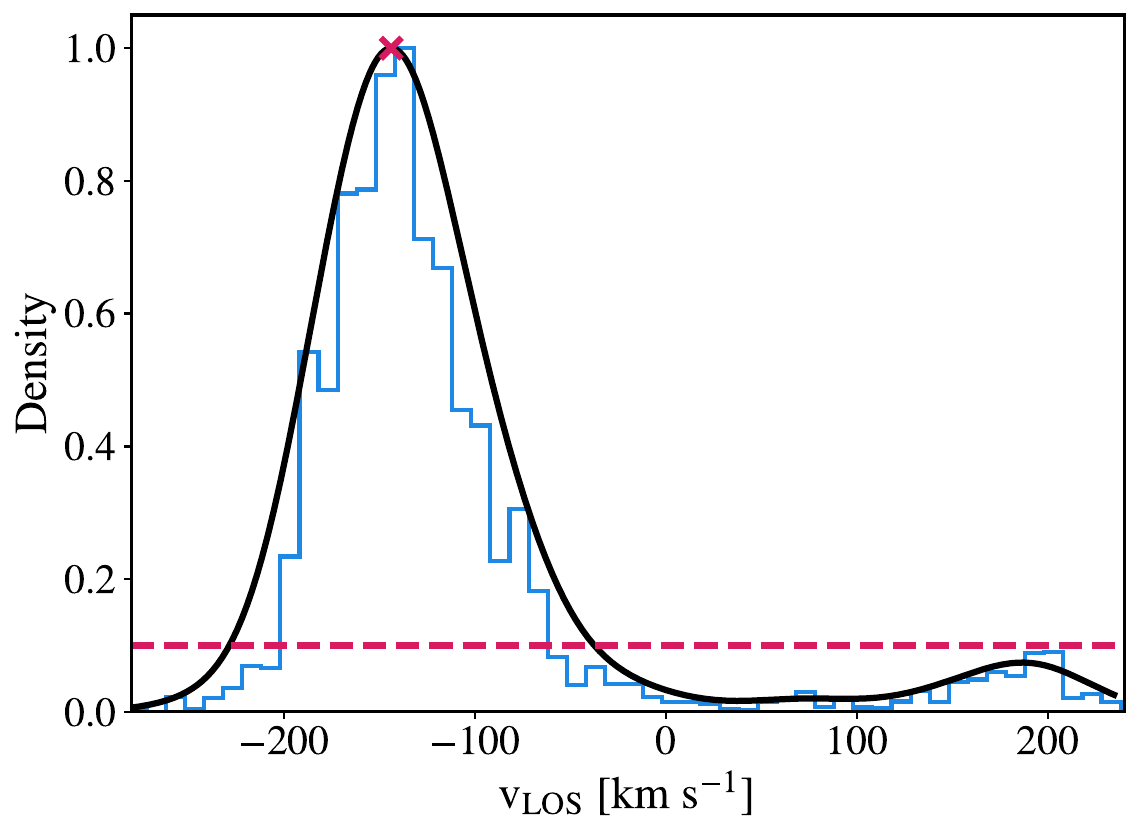}
        
    \end{subfigure}
    
    \vspace{0.1cm}
    
    \begin{subfigure}{0.49\linewidth}
        \centering
        \includegraphics[width=\linewidth]{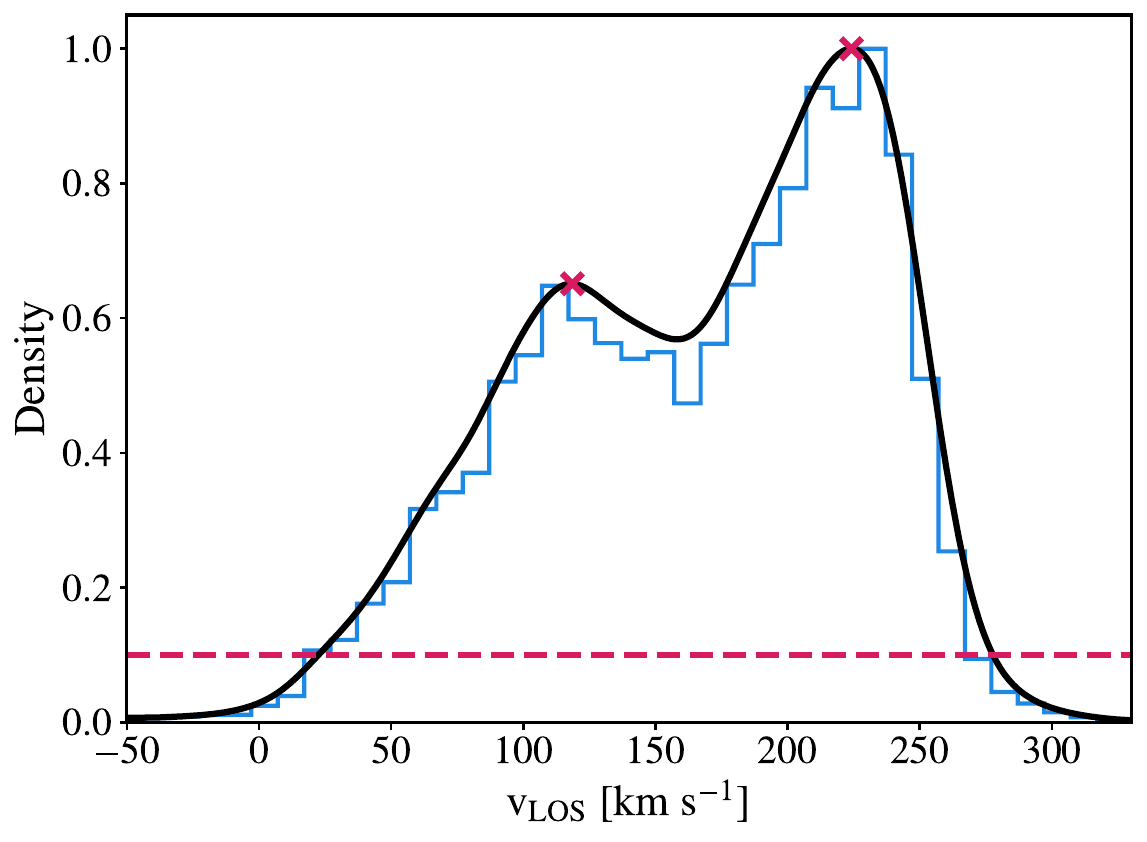}
        
    \end{subfigure}
    \hfill
    \begin{subfigure}{0.49\linewidth}
        \centering
        \includegraphics[width=\linewidth]{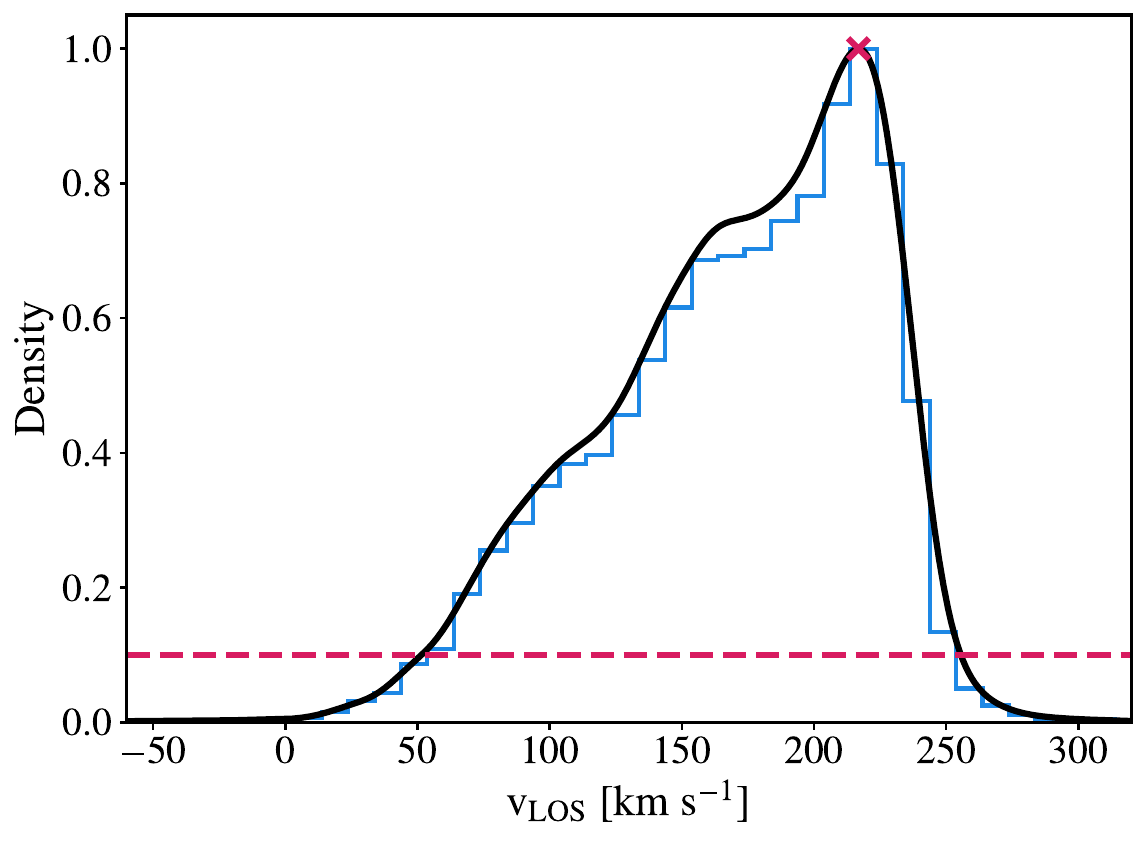}
    
    \end{subfigure}
    
    \caption{Examples of LOSVDs with their KDE representation and the peaks identified (red crosses). The KDEs (black solid lines) are compared to LOSVDs  binned to $10$ km s$^{-1}$ (blue step). The four panels illustrate how the peak detection algorithm works. The top panels show the effect of the minimum height parameter (highlighted by the red dashed line): a peak below this height is disregarded (see top right panel). The bottom panels illustrate how the prominence of the peaks is taken into account: only a single peak is detected in the bottom right panel.}
    \label{fig:peak_detection}
\end{figure*}
In this paper, we investigate the complexity of the LOSVDs extracted from Galaxy 102 (see Fig.\ref{fig:galaxy102} for a $g$-band face-on image of the galaxy, computed as explained in \citealp{Martig2012}). Galaxy 102 has a stellar mass of $4.3 \times 10^{10} M_{\odot} $ and an optical radius $R_{25}$ of $21.9$ kpc. A $2$D photometric decomposition with GALFIT \citep{Peng2010} was used to calculate a bulge fraction of $0.47$ \citep{Martig2012}. When viewed face-on Galaxy 102 has two main spiral arms. Both arms are somewhat fractured and appear to be very uneven in density. They are very tightly wound in the inner $10$ kpc of the galaxy and  become more open at larger radii. The arm that exists in the negative $x$ region appears to be denser and longer than the other arm.

The structure of Galaxy 102 is strongly affected by interactions. Its early disc is perturbed by a $1$:$15$ merger $8.5$ Gyr after the Big Bang, which resets the spiral structure and leaves the galaxy with a thickened and flared disc \citep{Marie2014}. After this, a thin disc forms and new spiral arms begin to grow. Their strength and shape is affected by a satellite orbiting around the disc for the last few Gyrs of the galaxy's life. This satellite has a nearly co-planar and co-rotating orbit and can be seen in Fig.\ref{fig:galaxy102}.

\subsection{LOSVD extraction and peak detection} \label{PD}

First, to align the galaxy edge-on, we compute the total angular momentum of all the stellar particles within the optical radius, and align it with the $z$ axis. The LOSVDs are then extracted as if the galaxy were viewed edge-on, projected in the $x-z$ plane. The LOSVD is then the distribution of the velocity of the star particles in the y direction, within a defined region. When extracting LOSVDs, one of the most important factors to consider is the number of particles that contribute to the histogram. 
We use the PowerBin package developed by \cite{Cappellari2025}  to perform adaptive spatial binning on the edge-on view of the galaxy, with an original grid size of $0.5$ kpc and a particle target number of $1000$ particles per adaptive bin. During the initial extraction of LOSVDs, it was found that a grid size smaller than $0.5$ kpc and fewer than $1000$ particles per bin resulted in  LOSVDs being too sensitive to small fluctuations.

A visual inspection of LOSVDs revealed that they often exhibit complex and asymmetrical shapes, sometimes with multiple peaks. We thus developed an automated way to detect those peaks firstly by smoothing the LOSVDs with a kernel density estimation (KDE) and then using a peak detection algorithm. 
The KDE algorithm from the SciPy library \citep{Virtanen2020} was applied to the stellar line-of-sight velocity values weighted by the luminosity of each particle in the g-band (as computed by  \citealp{Martig2012}). Each data point is represented by a Gaussian kernel, and the sum of each kernel produces an overall probability density function \citep{Silverman, wkeglarczyk2018kernel}. We use Scott's factor to determine the bandwidth of each kernel:
\begin{equation}
    h = \sigma n^{-\frac{1}{5}}
\end{equation}
where $h$ is the bandwidth, $\sigma$ is the standard deviation of the line-of-sight velocities, and $n$ is the number of data points \citep{scott}. This allows for the automatic optimisation of the bandwidth to ensure the underlying distribution is properly represented.  In Fig. \ref{fig:peak_detection}, we compare the KDE to histograms with a bin width equal to $10$ km\ s$^{-1}$ to confirm that the KDE is performing as expected.

Once the KDE has been applied, the find\_peaks function from the scipy.signal library is used to automatically detect peaks in the smoothed LOSVDs. This allows for the detection of local maxima by comparing the neighbouring values. 
Conditions may also be set to optimise peak finding. The first condition is on the minimum height of the peaks, represented by the pink dashed line in Fig. \ref{fig:peak_detection}. This height was set at $10\%$ of the tallest peak: any peak below this height was disregarded as it was deemed insignificant or noisy (see an example in the top right panel of Fig. \ref{fig:peak_detection}).
A prominence criterion was also applied,  which measures the height of the peak relative to its closest minimum. Any peak with a prominence below $0.01$ was disregarded to prevent small fluctuations in the LOSVD from being misidentified as genuine features. In Fig.\ref{fig:peak_detection}, the bottom-left panel shows a situation in which the smaller peak is retained because it has a prominence factor larger than $0.01$, whereas in the bottom-right panel the small fluctuation in the LOSVD is disregarded as a peak because it has a prominence factor that is too low.

This procedure allows us to robustly identify and count peaks in LOSVDs across the galaxy.

\section{Results}
In this section, we present the regions in which complex LOSVDs are present within Galaxy 102, discusses the origin of the multiple velocity peaks we observe, and shows how this impacts the values of $h_4$ measured throughout the disc.

\subsection{Spatial distribution of multiple-peaked LOSVDs}
\begin{figure}

    \begin{subfigure}{\linewidth}

        \includegraphics[width=\linewidth, trim= 0 45 0 0cm, clip]{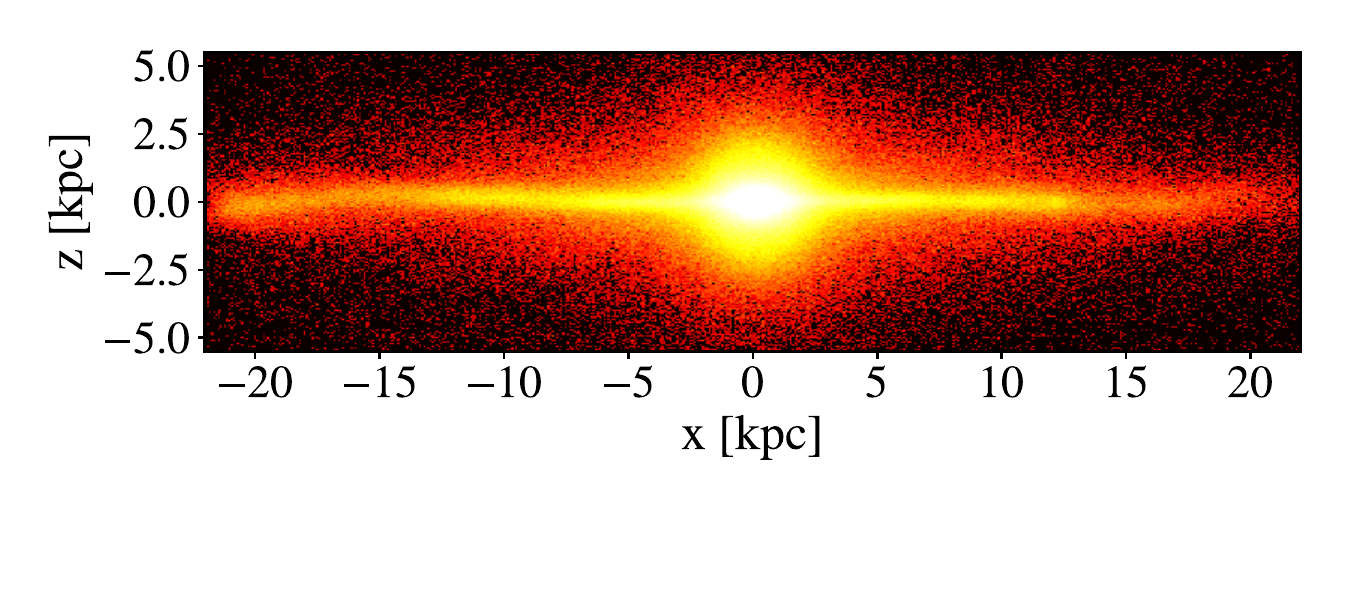}
    \end{subfigure}
    
    \begin{subfigure}{\linewidth}

        \includegraphics[width=\linewidth]{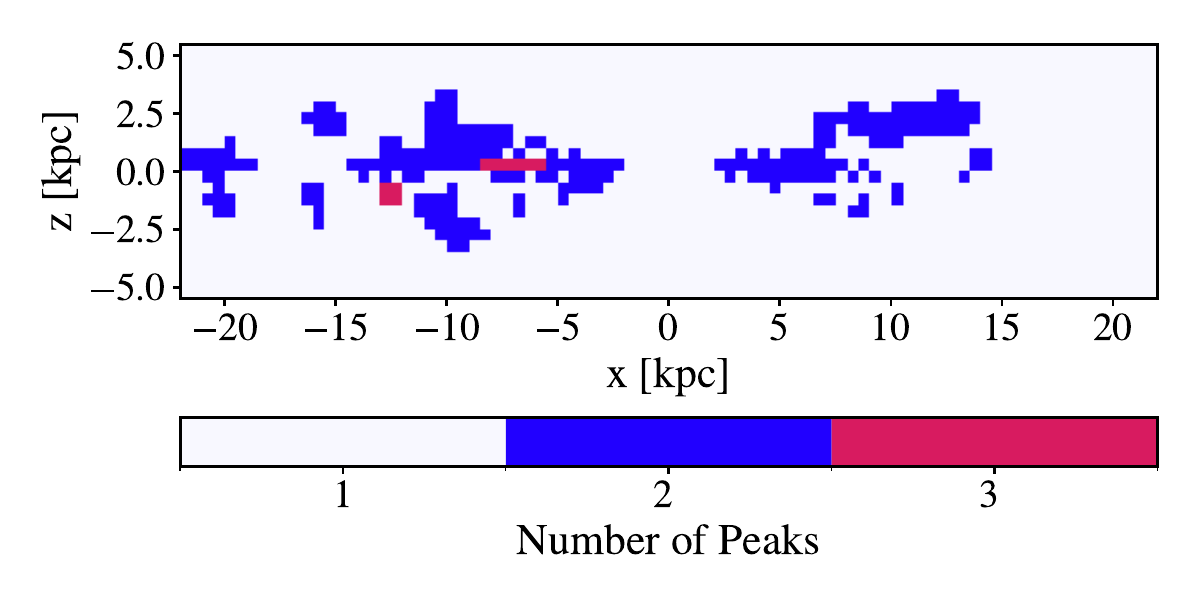}
    \end{subfigure}

    \begin{subfigure}{\linewidth}
      
        \includegraphics[width=1.03\linewidth]{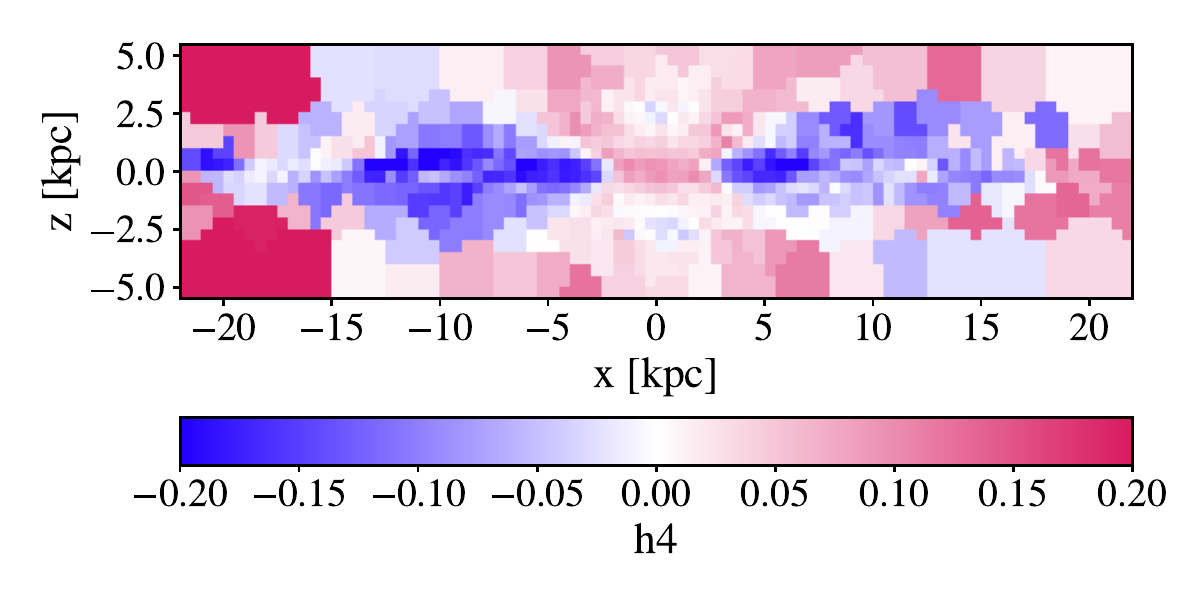}
    \end{subfigure}

    \caption{The top panel shows the luminosity-weighted, edge-on projection of Galaxy 102. The middle panel shows the number of peaks in each LOSVD across the disc, where multiple-peaked regions are mostly close to the midplane. The bottom panel shows the $h_4$ values across the disc; low $h_4$ regions trace multiple-peaked LOSVDs.}
    \label{fig:galaxy102_map}

\end{figure}
\begin{figure*}
    \centering

    \begin{subfigure}{0.49\linewidth}
        \hspace*{-1.5cm}
        \centering
        \includegraphics[height=7cm, keepaspectratio]{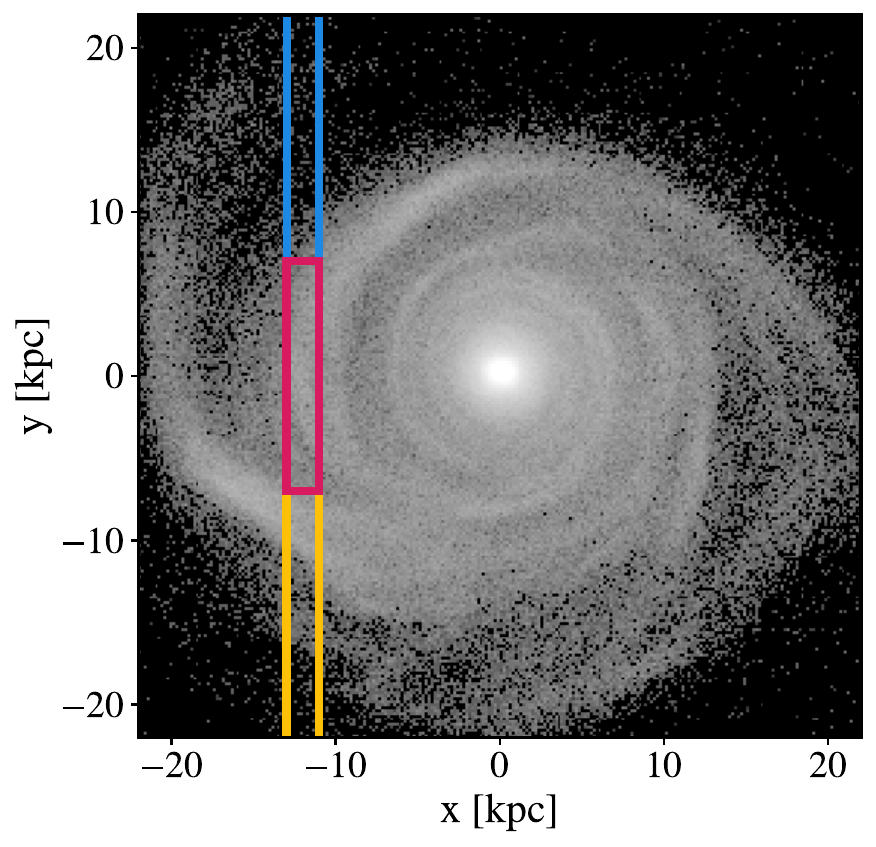}
    \end{subfigure}
    \hfill
    \begin{subfigure}{0.49\linewidth}
    \hspace*{-1cm}
        \centering
        \includegraphics[height=7cm, keepaspectratio]{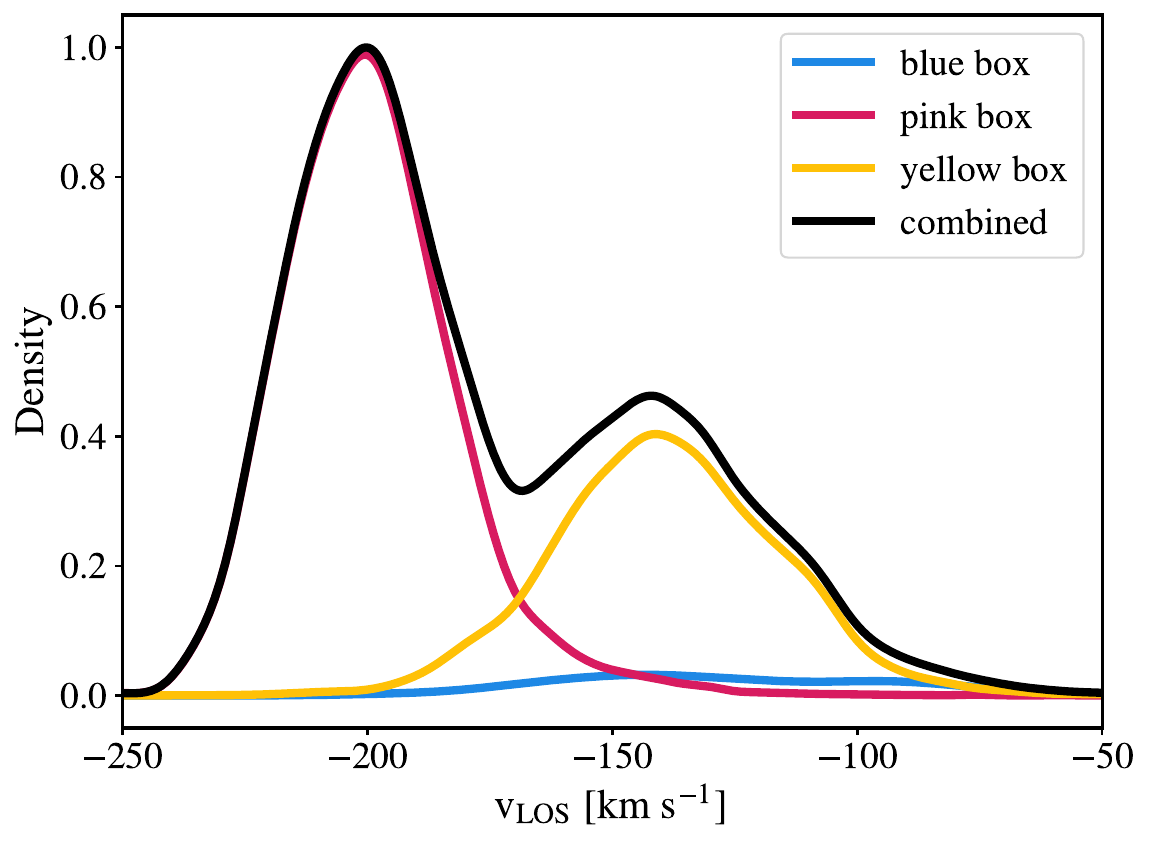}
    \end{subfigure}

    \begin{subfigure}{0.49\linewidth}
    \hspace*{-1.5cm}
        \centering
        \includegraphics[height=7cm]{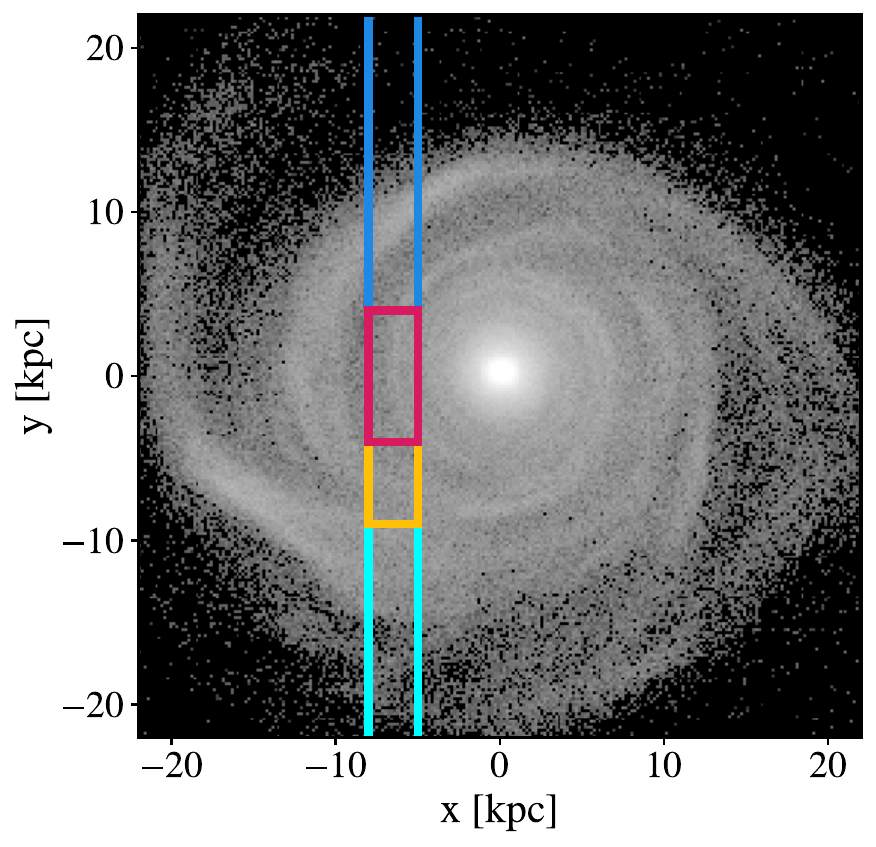}
    \end{subfigure}
    \hfill
    \begin{subfigure}{0.49\linewidth}
    \hspace*{-1cm}
        \centering
        \includegraphics[height=7cm]{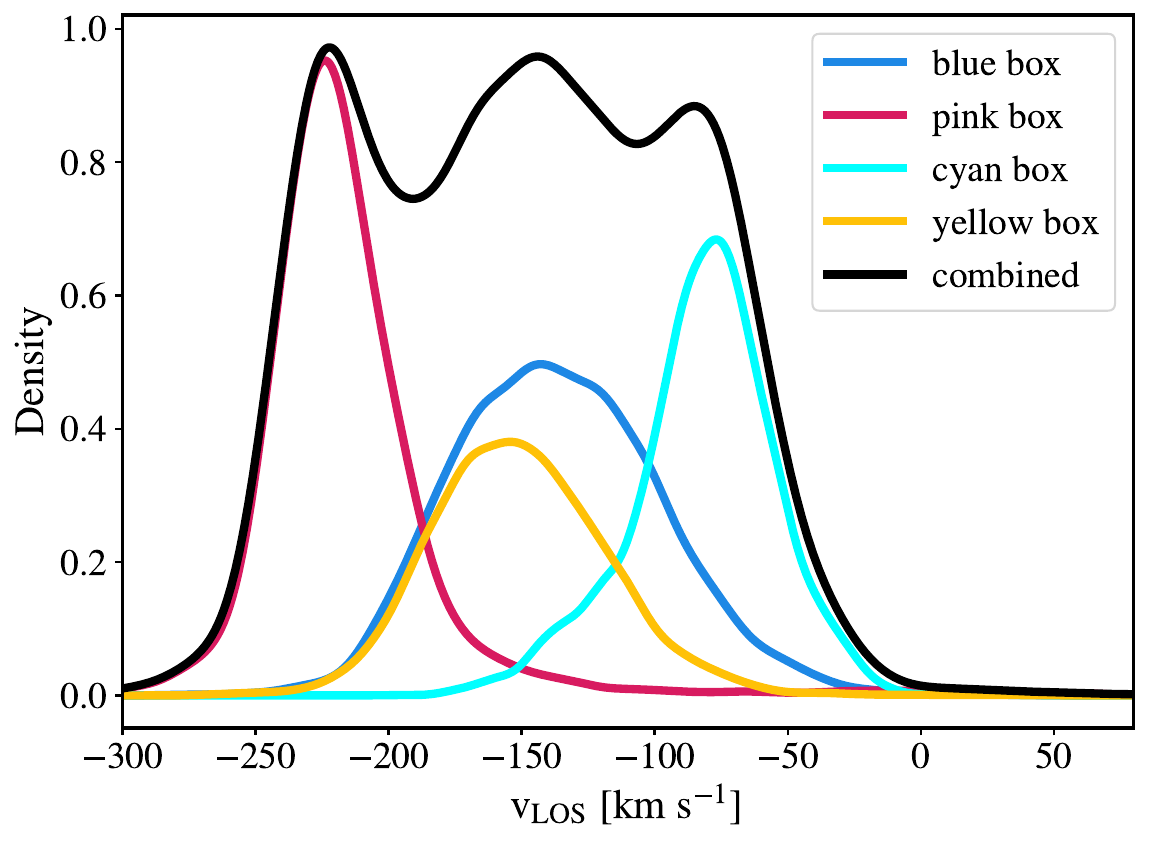}
    \end{subfigure}
    \caption{Illustration of how different regions of the disc contribute to LOSVDs. The top row is focused on a region of the galaxy with a double-peaked LOSVD. The left panel shows the face-on view of the galaxy, and three different boxes from which we extract LOSVDs. The top right panel shows those three LOSVDs, to illustrate how each region contributes to the total LOSVD. We find that the two peaks correspond to the central region of the galaxy and to the spiral in the outer disc. The bottom row shows a similar result, but for a region where the LOSVD has three peaks, where we find that peaks correspond to different spiral arms.}
    \label{fig:galaxy102_combined}  
\end{figure*}
We use the procedure described in Section 2.3 to extract LOSVDs from an edge-on view of Galaxy 102 and to count the number of peaks in those LOSVDs.
We limit our analysis to particles with $|x|< R_{25}$, $|y|< R_{25}$ and $|z|< 5$ kpc (see top panel of Fig. \ref{fig:galaxy102_map}). This excludes the satellite currently orbiting the galaxy.
In the middle panel of Fig. \ref{fig:galaxy102_map}, we show a map of the number of peaks in LOSVDs across the galaxy. Galaxy 102 exhibits multiple-peaked LOSVDs throughout its disc and these multiple-peaked LOSVDs are most prominent along the midplane. In most cases, we observe double-peaked LOSVDs, with a few examples of triple-peaked LOSVDs.

The very centre of the galaxy has simple LOSVDs, and  regions with multiple-peaked LOSVDs begin $\approx \pm 2$ kpc from the centre and extend as far out as $R_{25}$ in the negative $x$ side. Up to $x\approx \pm 6$ kpc, the complex regions are confined to $|z| <1 $ kpc, and also exhibit a somewhat symmetric structure. Towards the edge of the disc, the regions with double-peaked LOSVDs become more patchy and can be found at larger heights. The two regions with triple-peaked LOSVDs are only found very close to the midplane.
In the next Section, we explore the physical origin of those multiple peaks.

\subsection{Origin of multiple-peaked LOSVDs}

Given the active accretion history of Galaxy 102, and the current interaction with a satellite, it could have been expected that some of the peaks in LOSVDs would be due to accreted stars or gas, creating complex kinematic components in this galaxy. However, we find that it is not the case, and none of the multiple peaks shown in Fig. \ref{fig:galaxy102_map} are due to interactions. Instead, they reflect the presence of strong spiral arms with the disc.

To show this, we first focus on a region with 
a double-peaked LOSVD,  at $-13<x<-11$ kpc and  $0<z<0.5$ kpc.
The top-left panel of Fig. \ref{fig:galaxy102_combined} shows the face-on view of Galaxy 102 and the top right panel shows the corresponding LOSVD. We split the line-of-sight into three separate boxes, and plot their individual contributions to the LOSVD in the top right panel of Fig. \ref{fig:galaxy102_combined}. 
The pink region encapsulates the line of nodes and is the main contributor to the fastest and tallest peak. This is expected for a circular orbit as the stellar motion is directly aligned with the line-of-sight, resulting in a large line-of-sight velocity. The density is also expected to be larger, as the inner regions of a galaxy are denser than the outer regions. The blue and yellow regions are very different from the pink region. The blue region does not contain many stars, making its contribution very low. By contrast, the yellow region makes up the majority of the second peak. Indeed, it contains a portion of a spiral arm, and the density of the spiral arm is significant enough to produce a separate peak in the distribution. The line-of-sight velocity value of the second peak is lower than that of the central regions, as the projection of the velocity vector on the line-of-sight  is lower in the outer regions of the galaxy.

We similarly study the origin of peaks in the triple-peaked LOSVD extracted from the region defined by $-8<x<-5$, $0<z<0.5$ kpc (see bottom row of Fig. \ref{fig:galaxy102_combined}). This time, we split the face-on view into four different components. The pink region is the main contributor to the fastest and tallest peak, as expected. The blue and yellow regions contain a section of the same spiral arm, and together they contribute to the middle peak. As they are at a similar radius from the galactic centre, and because the spiral arm is quite uniform in density, those two regions have very similar contributions to the global LOSVD.

The cyan region contributes mainly to the slowest peak. This is because the stars that exist in the cyan box are moving almost perpendicularly to the line-of-sight. Hence, the presence of a spiral arm causes an overdensity within this region and produces a separate peak at low line-of-sight velocity.

We note that spiral arms could also have an imprint on LOSVDs by locally perturbing the kinematics of stars in the disc. However, spiral arms are usually associated with peculiar velocities of order of only 10 \kms \citep{2015Grand,2016Monari}: this effect is thus very small and  does not impact our results as the peaks we find have a separation of more than $\approx 50$ \kms. This shows that the multiple peaks we observe in the LOSVDs are due to the overdensities of stars that exist in spiral arms, and that variations in the number of arms present along a line-of-sight create variations in the number of peaks found in LOSVDs.
In the next Section, we explore how this might be used to study the strength of spiral arms across the disc of an edge-on galaxy.

\subsection{Variations in strength of spiral arms} \label{var}

\begin{figure}

    \hfill
    \begin{subfigure}{\linewidth}
        \centering
        \includegraphics[width=\linewidth]{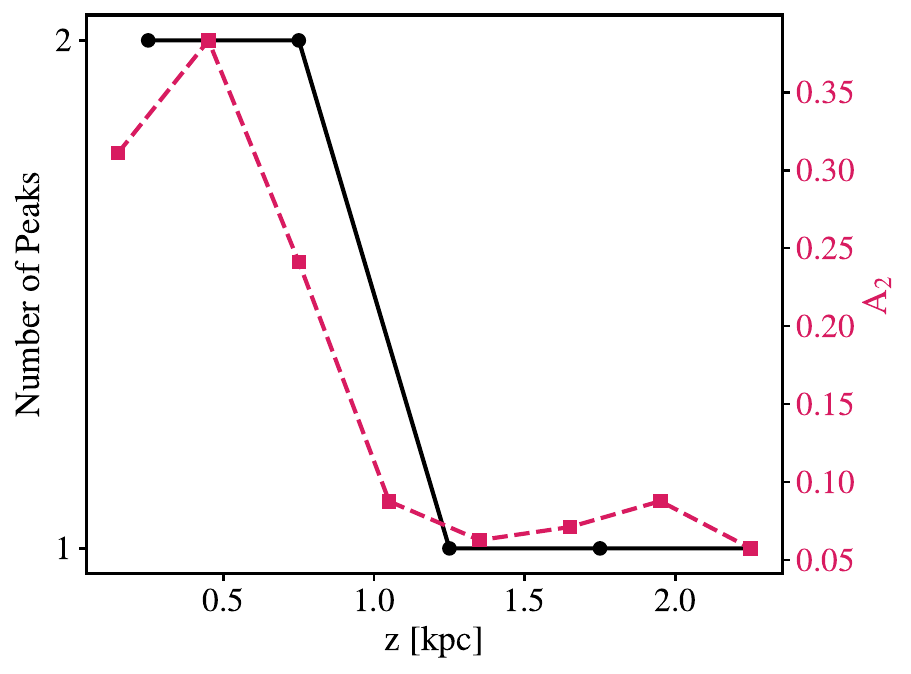}
    \end{subfigure}
    \caption{Number of peaks in LOSVDs (black solid line) and strength of spiral arms (pink dashed line) as a function of height above the midplane. Starting from the same position within the disc as for the double-peaked LOSVD shown in Fig.\ref{fig:galaxy102_combined}, we then iterate from $z=0$ to $z=2.5$ kpc in steps of 0.5 kpc, and compute the number of peaks as a function of $z$: the LOSVDs become single-peaked at large $z$. We compare this with how the $A_2$ value changes as a function of $z$: it begins to plateau where the double-peaked LOSVD becomes single-peaked, indicating that this is where the spiral arm disappears.
}
    \label{fig:spiral_thickness}
\end{figure}

\begin{figure}
    \centering
    \begin{subfigure}{\linewidth}
        \centering
        \includegraphics[width=\linewidth]{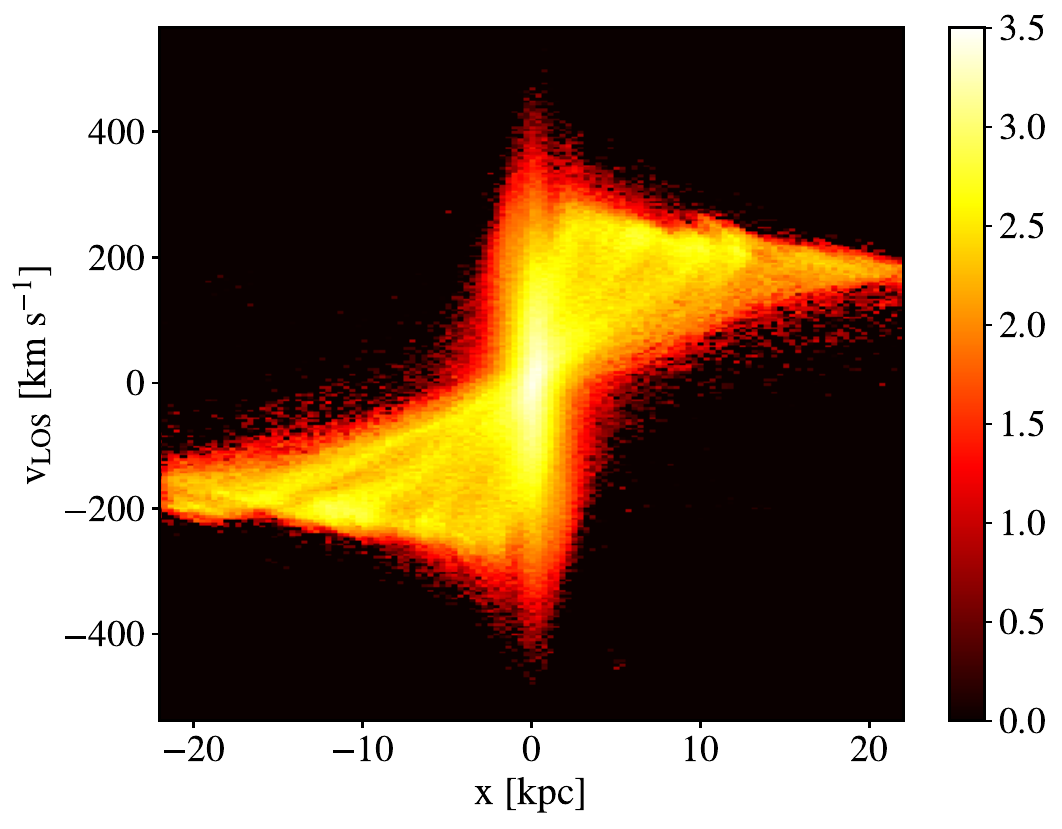}
    \end{subfigure}
    
    \begin{subfigure}{\linewidth}
        \centering
        \includegraphics[width=\linewidth]{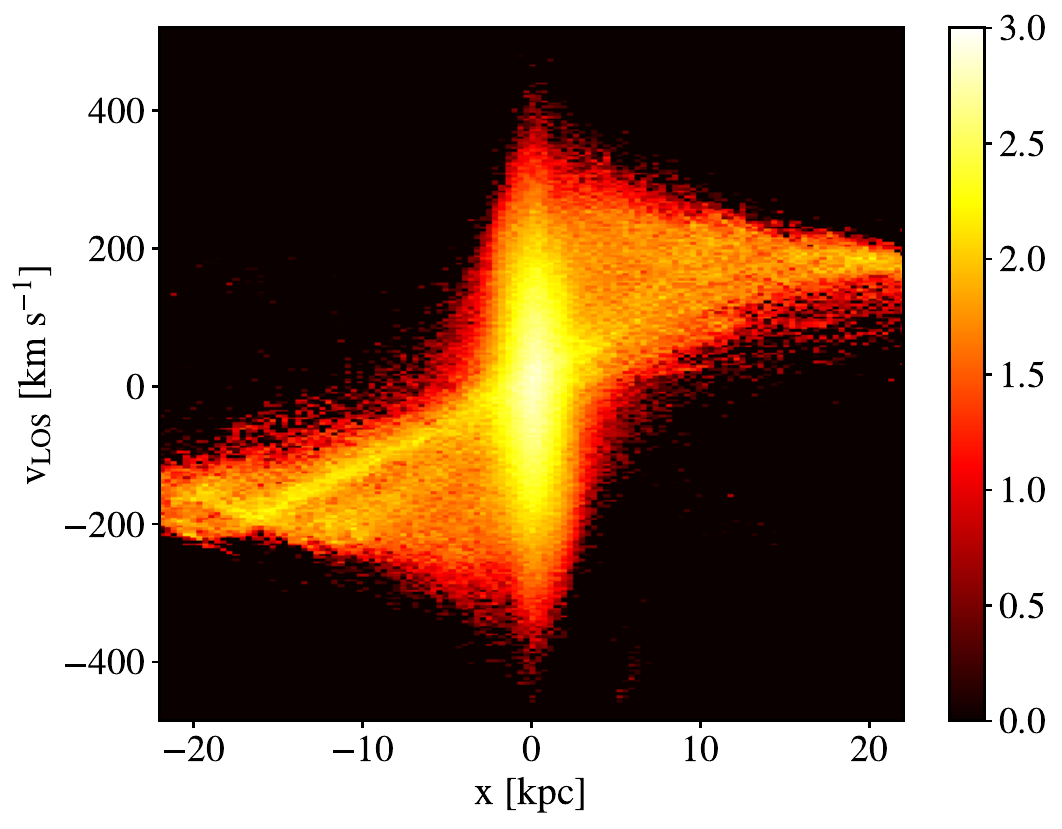}
    \end{subfigure}
    \caption{Position-Velocity Diagrams at two different heights: the top panel shows the PVD in the midplane ($0<z<0.5$ kpc) and the bottom panel shows the PVD above the midplane ($0.5<z<1.5$ kpc). The colour bar represents the number of stars in each bin, in logarithmic scale. The ridges arise due to spiral arms and only one, lower velocity ridge remains above the midplane. This indicates that the spiral arm that exists at a large galactocentric radius is thicker than the inner spiral arms.}
    \label{fig:PVD}
\end{figure}

As discussed in the previous section, the complexity observed in LOSVDs might be used to trace spiral arms. In Fig. \ref{fig:galaxy102_map}, we notice that multiple-peaked LOSVDs are commonly found close to the midplane, and that they tend to become single-peaked at large heights.
In this Section, we explore if the transition from two peaks to one peak may be an indication of the decreasing strength of the spiral arms as a function of height above the midplane.

We investigate the vertical variations in LOSVDs taken from a region $-13<x<-11$ kpc (see top right panel of Fig.\ref{fig:galaxy102_combined}). 
We define vertical bins from $z=0$ to $z =2.5$ kpc with a step size of $0.5$ kpc, and show in Fig. \ref{fig:spiral_thickness} how the number of peaks in LOSVDs varies as a function of $z$.
Close to the midplane, the LOSVDs are double-peaked, as seen in the top panel of Fig. \ref{fig:galaxy102_combined}. We then observe 
a transition from two peaks to one occurring at $z = 1.25$ kpc.

To determine whether this change is due to the spiral structure disappearing at large heights, we quantify the strength of spiral arms as a function of $z$. We compute the A$_2$ Fourier mode of the same spiral structure, calculated using:
\begin{equation}
    A_2 =\frac{1}{N}\sqrt{\left(\sum_i{\cos\theta_i}\right)^2 +\left(\sum_i {\sin\theta_i}\right)^2}
\end{equation}
where $N$ is the number of stars per annulus and $\theta$ is the angular position of the stars. The A$_2$ value is calculated within an annulus with an inner radius of $13$ kpc and an outer radius of $17$ kpc. This encloses the part of the spiral arm that contributes to the second peak in the LOSVD, as seen in the yellow box in the top panel of Fig. \ref{fig:galaxy102_combined}. The A$_2$ value is calculated from $z=0$ to $z=2.5$ kpc with a step size of $0.25$ kpc (see Fig. \ref{fig:spiral_thickness}). In the midplane, A$_2$ is close to 0.5, indicating a strong two-armed spiral structure (A$_2$ $> 0.1$ is typical of a strong spiral structure, see e.g. \citealp{2011Fujii, 2014Dobbs}). Then, A$_2$ decreases until just after $z =1.25$ kpc, where it begins to plateau to a value below 0.1. This indicates that the strength of the spiral arm and the number of peaks in LOSVDs both decrease at the same height. Hence, multiple-peaked LOSVDs may be a useful tool in determining the thickness of spiral arms in edge-on galaxies.
\begin{figure}
    \centering
    \includegraphics[width=\linewidth]{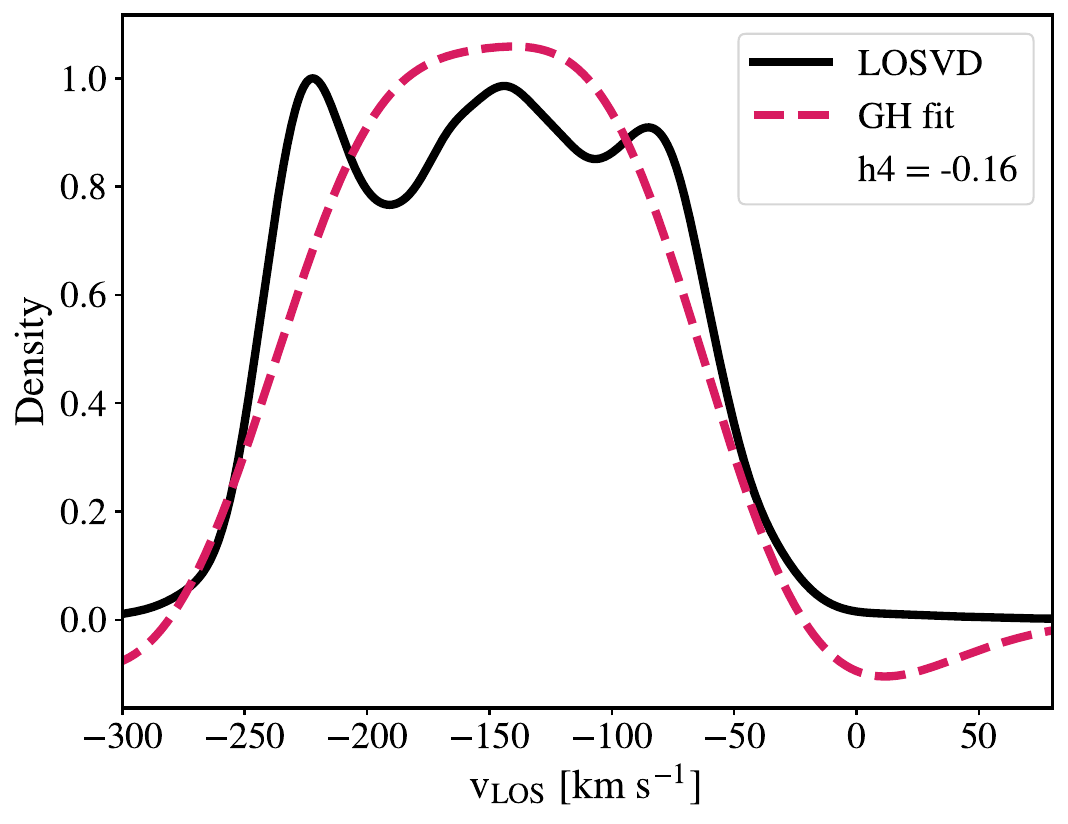}
    \caption{Gauss-Hermite polynomial fit (pink dashed line) to the triple-peaked LOSVD from the bottom right panel in Fig. \ref{fig:galaxy102_combined} (black solid line). This corresponds to $h_4=-0.16$: this negative value arises due to multiple peaks occurring over a small velocity range.}
    \label{fig:$h_4$_fit}
\end{figure}

As can be seen in Fig. \ref{fig:galaxy102_map}, the multiple-peaked regions are the thickest at larger radii. This indicates that the spiral arms that exist at a larger galactocentric radius are thicker than the spiral arms in the inner disc.
To further investigate this, we extract Position-Velocity Diagrams (PVDs) both in the midplane ($0 - 0.5$ kpc) and above the midplane ($0.5-1.5$ kpc), as shown in Fig. \ref{fig:PVD}). In the midplane, the PVD shows multiple ridges which correspond to different sections of spiral arms (see also Fig. 10 in \citealp{2013Davis} for signatures of spiral arms in molecular gas tracers in PVDs). First we investigate the left hand side of the PVDs. The ridge that occurs at the lowest velocities corresponds to the section of the arm found in the outer disc (crossing the yellow box in the top panel of Fig. \ref{fig:galaxy102_combined}).
Above the midplane, only this ridge in the PVD remains throughout the disc and the ridges corresponding to spiral arms found at smaller galactocentric radii disappear. Next we investigate the right hand side of the PVDs. Close to the midplane the PVDs are dominated by stars on the line of nodes, meaning that the spiral arms are too insignificant in density compared to the main ridge to produce a prominent second ridge. As the distance above the midplane increases, the density on the line of nodes and in the main spiral arm become more similar. This explains why, in Fig. \ref{fig:galaxy102_map}, the number of peaks in LOSVDs increases from one to two when moving above the midplane between $x = 10$ and $x = 15$ kpc. 

The presence of low velocity ridges in the PVD above the midplane indicates that the spiral arms that exist at large galactocentric radii are thicker than in the inner galaxy.
This could simply be due to the general increase of disc thickness as a function of radius: \cite{Marie2014} have shown that the disc of Galaxy 102 is flared. However, it is also possible that the outer spiral arm could have a different origin from the spiral arms in the inner disc (it could be excited by the satellite orbiting the galaxy), and that this could result in different thicknesses.

\subsection{Gauss-Hermite parameters}
Reconstructing full LOSVDs from spectra is not commonly done, and most studies on galaxy kinematics parametrise LOSVDs using Gauss-Hermite coefficients. We thus fit LOSVDs in each spatial bin with a 4$^\mathrm{th}$ order Gauss-Hermite polynomial to study the impact of spiral arms on $h_3$ and $h_4$. In a simple axisymmetric disc viewed edge-on, we expect $h_3$ to be anti-correlated with $V$, and $h_4$ to be positive in the outer disc \citep{van_der_Marel, 1993Gerhard, Kregel2005, wang2024}.

Instead, we find that in Galaxy 102 the $h_3$ map is very complex (see Fig. \ref{fig:GHmaps}). A general anti-correlation between $V$ and $h_3$ is seen in the outer disc and at large heights, but many other regions deviate from this simple pattern. The $h_3$ map is hard to interpret as a result of the presence of spiral arms. The value of $h_3$ changes based on the density contrast between the peaks, with the more prominent peak determining which way the LOSVD is skewed. As discussed in Section \ref{var}, the peaks can shift in relative contribution throughout the disc, meaning the $h_3$ values can also shift dramatically.
In contrast, the $h_4$ map (see Fig. \ref{fig:galaxy102_map}) shows negative regions where LOSVDs have multiple peaks.
To illustrate why this is the case we show in Fig. \ref{fig:$h_4$_fit} the triple-peaked LOSVD from Fig. \ref{fig:galaxy102_combined} with its Gauss-Hermite fit, where $h_4$ reaches a low value of $-0.16$. Indeed, if an LOSVD has multiple peaks that occur over a relatively small velocity range, the LOSVD appears very flat-topped  and the value of $h_4$ will be negative. Therefore, the regions of negative $h_4$ in Galaxy 102 arise as a result of multiple-peaked LOSVDs due to spiral arms. It can be seen from Fig.\ref{fig:galaxy102_map} that there are also negative values of $h_4$ in the regions surrounding the multiple-peaked region. These are regions in which the over densities do not meet the peak-detection criteria, but still influence how flat-topped the LOSVDs are. 

It is worth noting that there is no one-to-one association between  negative $h_4$ and multiple-peaked LOSVDs. On the one hand, negative $h_4$ values can  arise even in certain areas of axisymmetric discs \citep{Kregel2005}. On the other hand, double-peaked LOSVDs can correspond to positive values of $h_4$ if the second peak occurs far from the main peak (see for instance regions of positive $h_4$ values in UGC 00903, a galaxy with two counter-rotating discs, \citealp{Fraser_McKelvie_2025}). Multiple-peaked regions only correspond to a negative $h_4$ when the peaks are relatively close in velocity, which is the case for peaks produced by multiple spiral arms.

\section{Observability of spiral arm signatures}
So far we have shown how to use multiple-peaked LOSVDs and regions of negative $h_4$ to detect the presence of spiral arms in a simulated edge-on disc galaxy. The first step of that analysis involved applying a KDE to the distributions of line-of-sight velocities. The smoothness of each KDE is governed by a bandwidth, which we computed using Scott's factor (see section \ref{PD}): this results in different bandwidths for each bin, with values between 4 and 29 km s$^{-1}$, and an average of 16 \kms.

However, in order to assess the observability of spiral arm signatures using current spectrographs, we need all LOSVDs to have the same bandwidth, which should be equal to the spectrographs' velocity resolution. We thus convolve each LOSVD with a Gaussian of fixed width, and explore how much information is retained when we decrease the velocity resolution. We consider velocity resolutions varying from 20 km s$^{-1}$, which is achievable with WEAVE \citep{2024Jin} and KOALA \citep{2012Ellis} in high resolution mode, to 50 km s$^{-1}$, achievable with MUSE \citep{2010Bacon}.

\begin{figure*}
\centering

\begin{minipage}{0.48\textwidth}
    \centering
    \includegraphics[width=\linewidth]{fig3b.pdf}
\end{minipage}\hfill
\begin{minipage}{0.48\textwidth}
    \centering
    \includegraphics[width=\linewidth]{fig3c.pdf}
\end{minipage}

\vspace{0.5cm}

\begin{minipage}{0.48\textwidth}
    \centering
    \includegraphics[width=\linewidth]{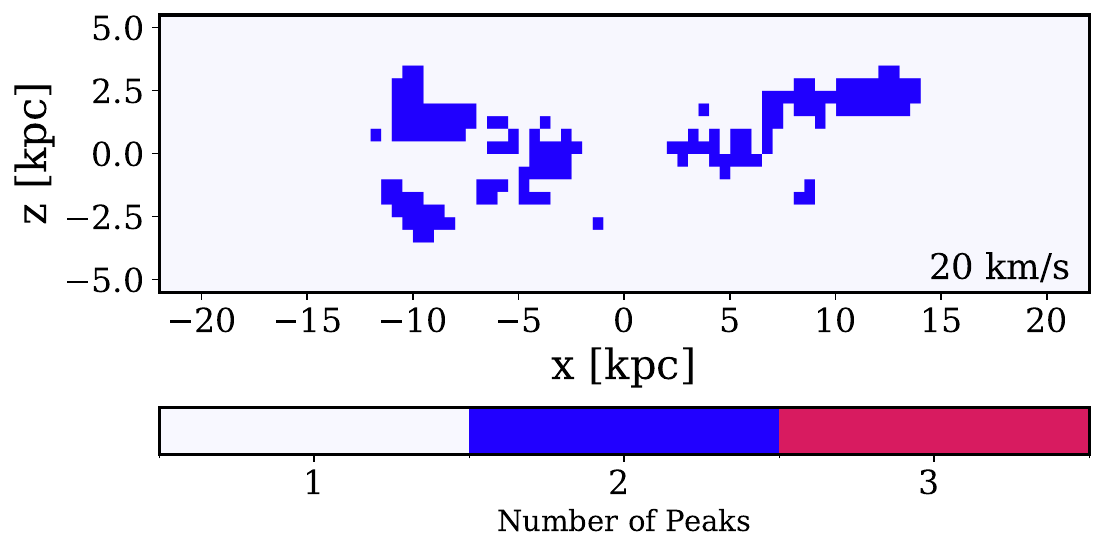}
\end{minipage}\hfill
\begin{minipage}{0.48\textwidth}
    \centering
    \includegraphics[width=\linewidth]{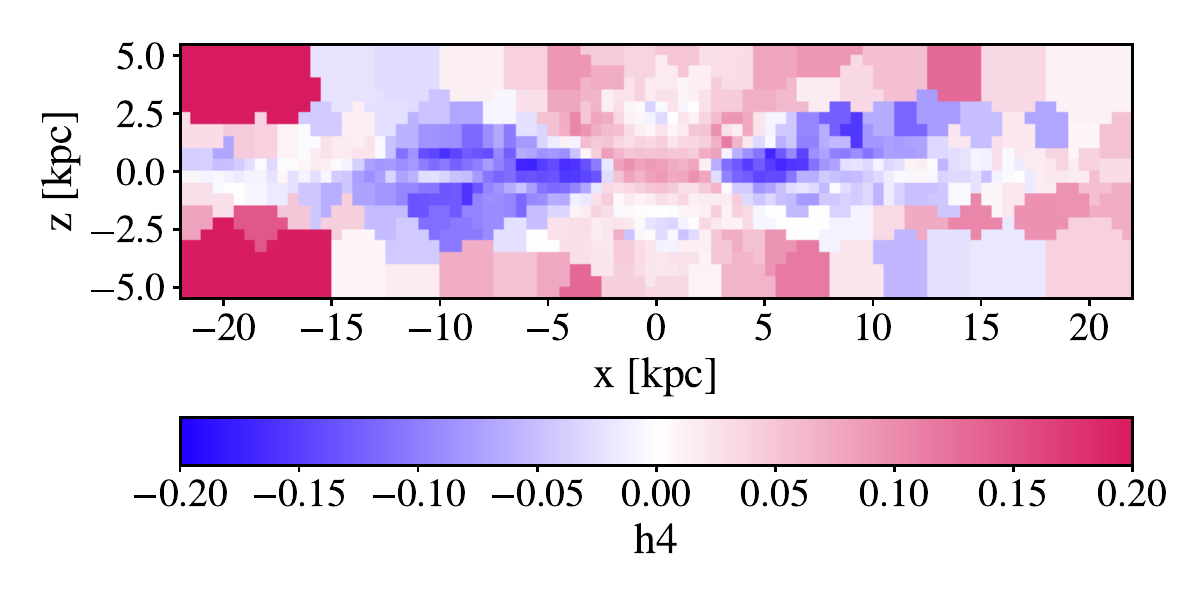}
\end{minipage}

\vspace{0.5cm}

\begin{minipage}{0.48\textwidth}
    \centering
    \includegraphics[width=\linewidth]{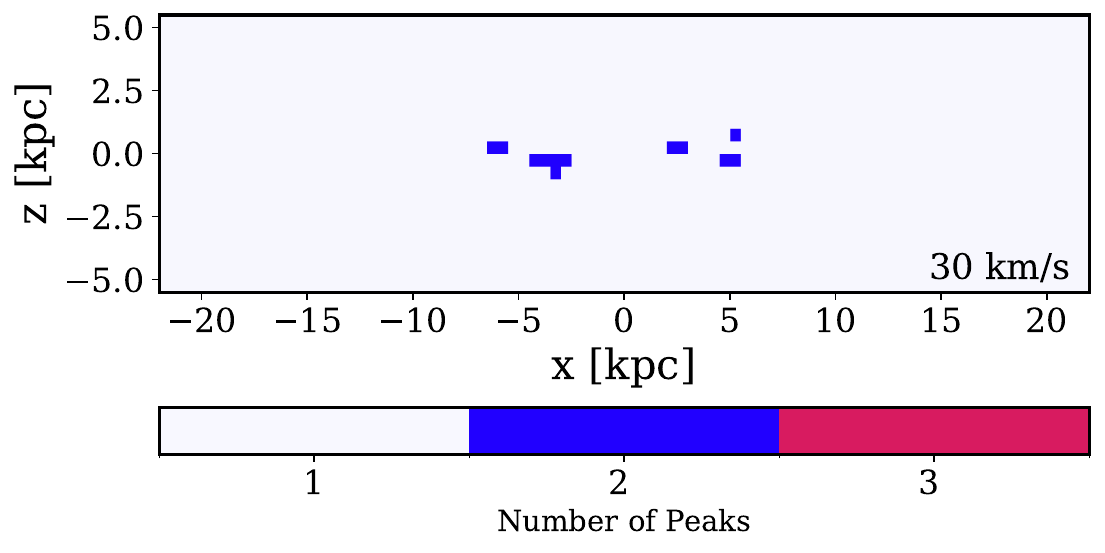}
\end{minipage}\hfill
\begin{minipage}{0.48\textwidth}
    \centering
    \includegraphics[width=\linewidth]{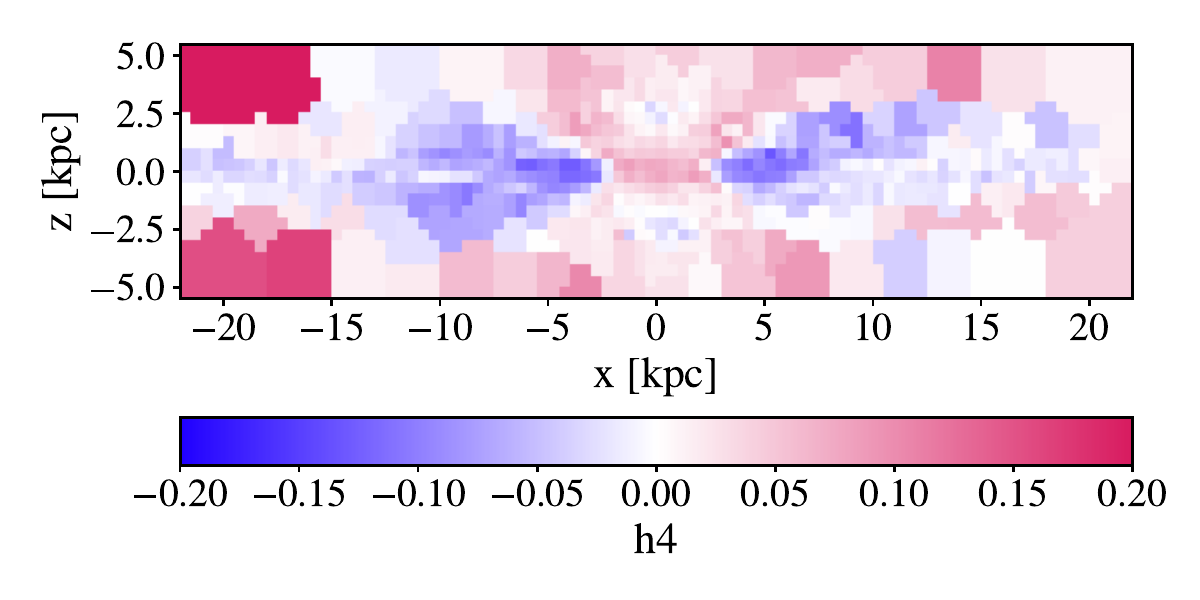}
\end{minipage}

\vspace{0.5cm}

\begin{minipage}{0.48\textwidth}
    \centering
    \includegraphics[width=\linewidth]{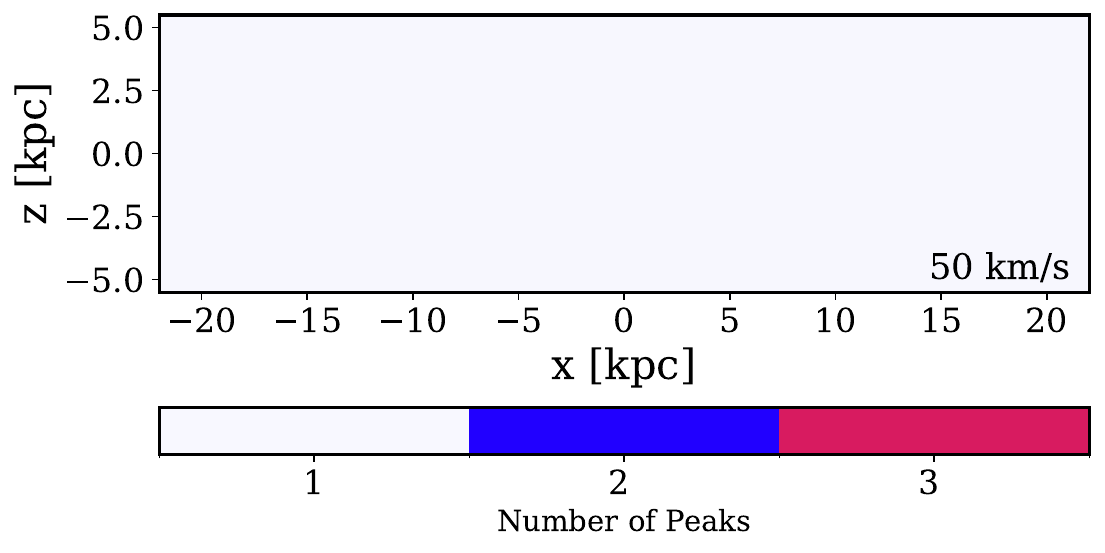}
\end{minipage}\hfill
\begin{minipage}{0.48\textwidth}
    \centering
    \includegraphics[width=\linewidth]{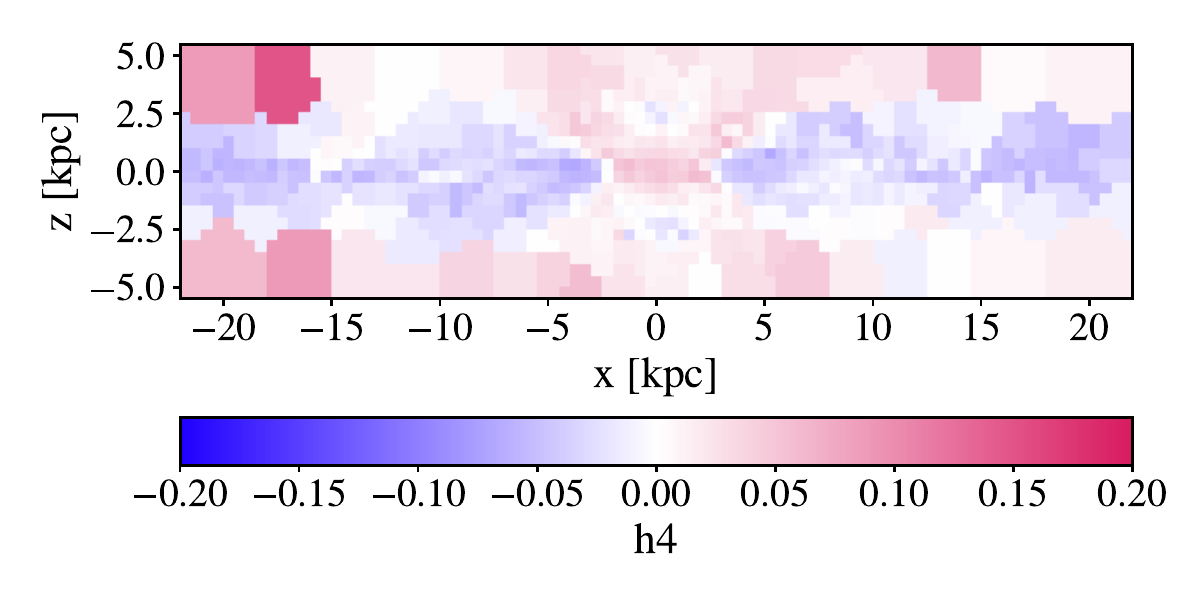}
\end{minipage}

\caption{Effect of velocity resolution on maps of the number of peaks (left panels) and $h_4$ (right panels). The top row shows the fiducial setup used throughout the paper, and the next rows show results for velocity resolution degraded to 20, 30 and 50 \kms. The number of peaks we can resolve decreases as the velocity resolution decreases, and all multiple-peaked regions disappear beyond a velocity resolution of $30$ \kms. The $h_4$ maps retain their shape as the velocity resolution decreases, the only noticeable change is all values move closer to $0$. }
\label{fig: resoltion difference}
\end{figure*}

\begin{figure*}
    \centering
    
    \begin{subfigure}{0.49\linewidth}
        \centering
        \includegraphics[width=\linewidth]{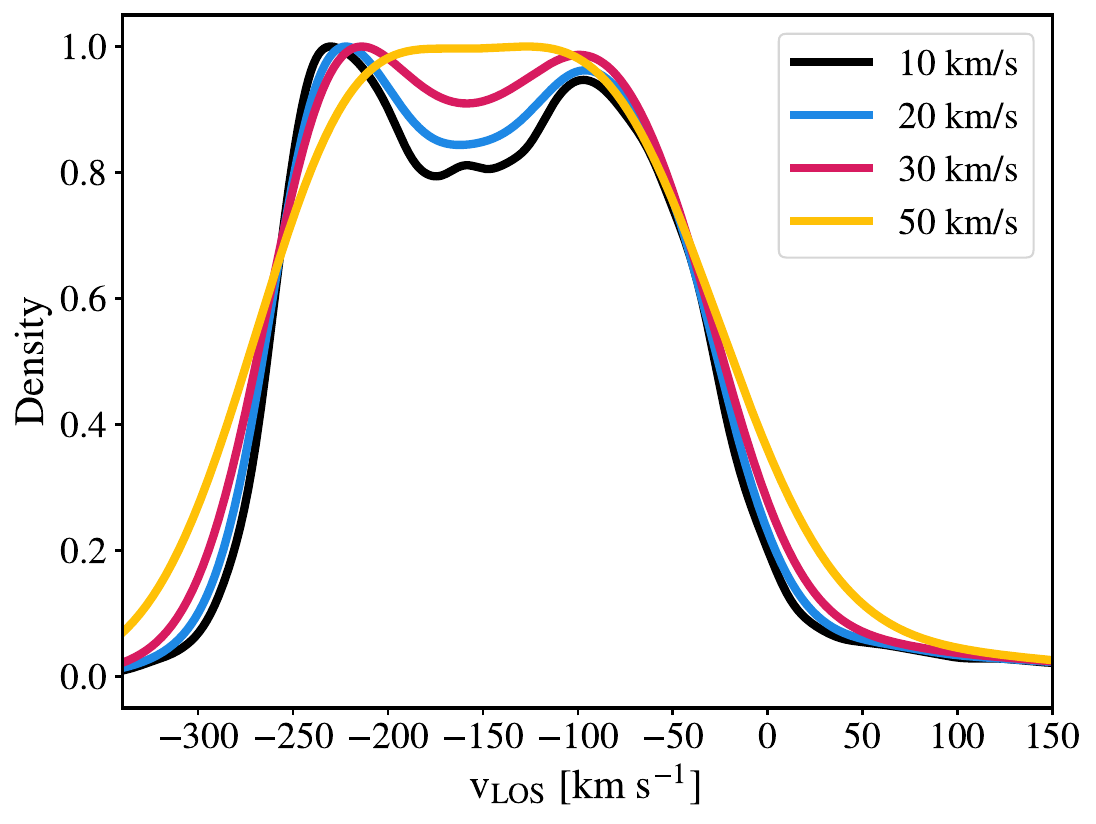}
       
    \end{subfigure}
    \hfill
    \begin{subfigure}{0.49\linewidth}
        \centering
        \includegraphics[width=\linewidth]{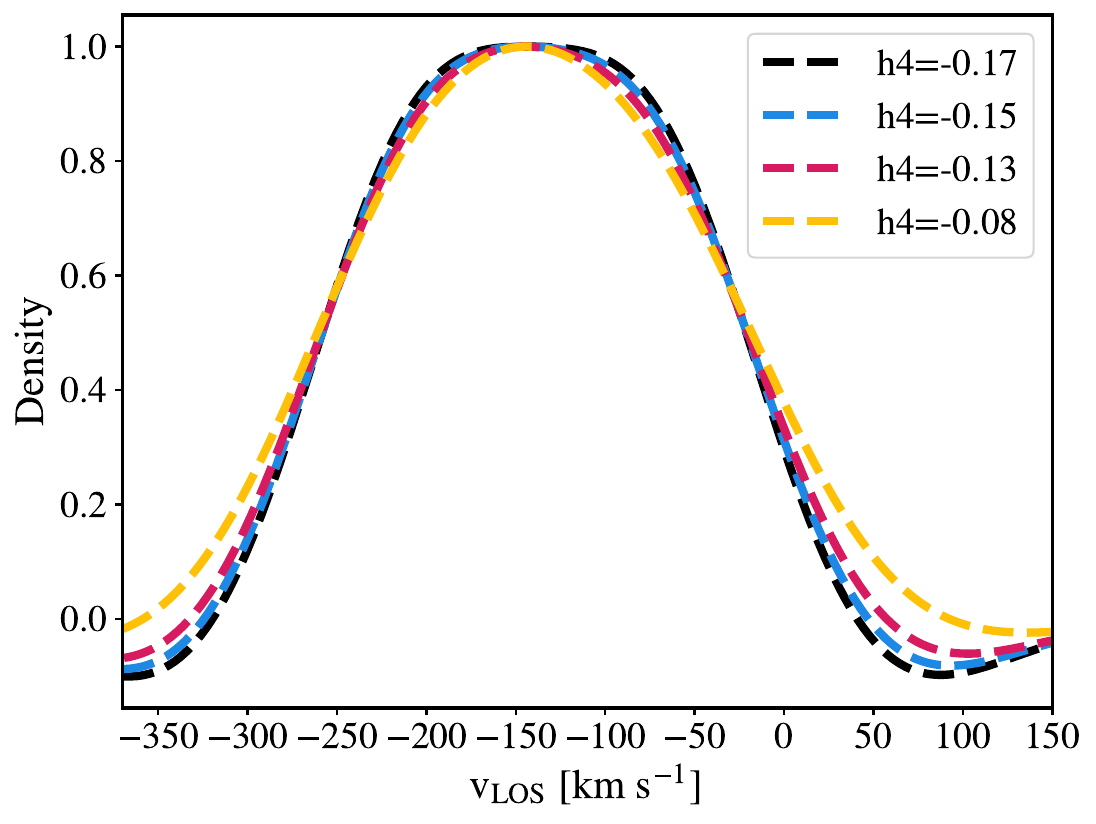}
        
    \end{subfigure}
    
    \vspace{0.1cm}
    
    \begin{subfigure}{0.49\linewidth}
        \centering
        \includegraphics[width=\linewidth]{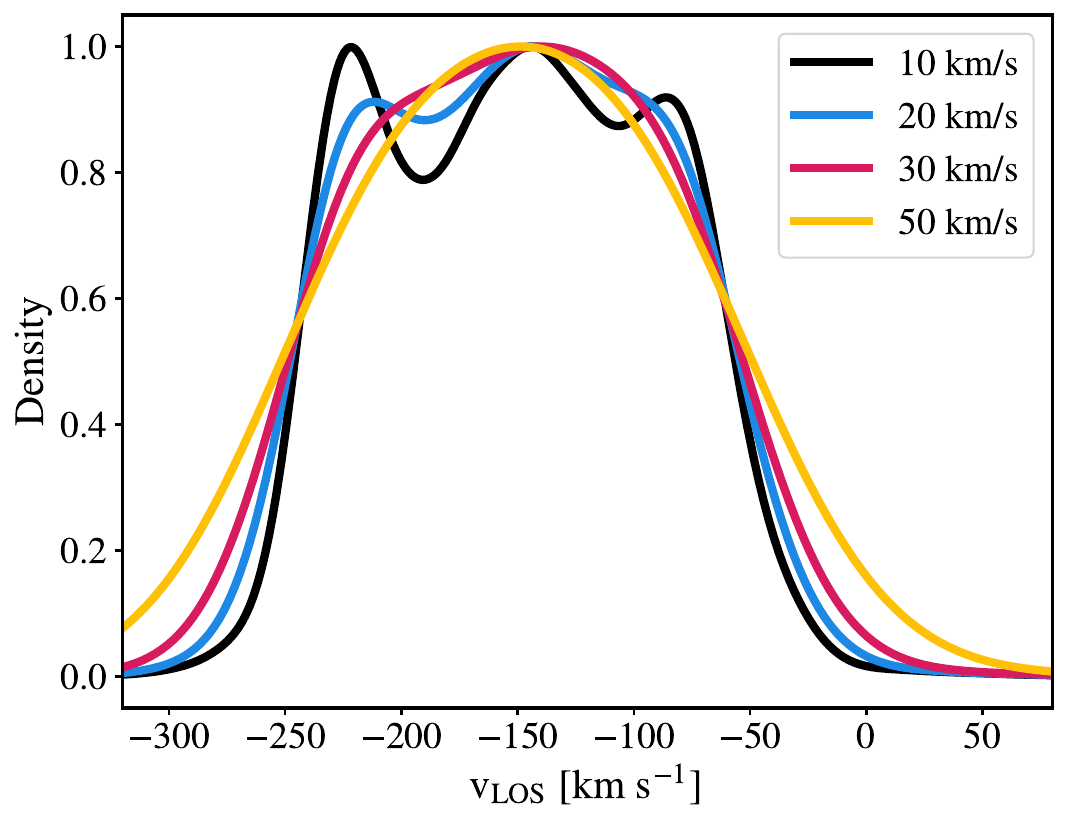}
        
    \end{subfigure}
    \hfill
    \begin{subfigure}{0.49\linewidth}
        \centering
        \includegraphics[width=\linewidth]{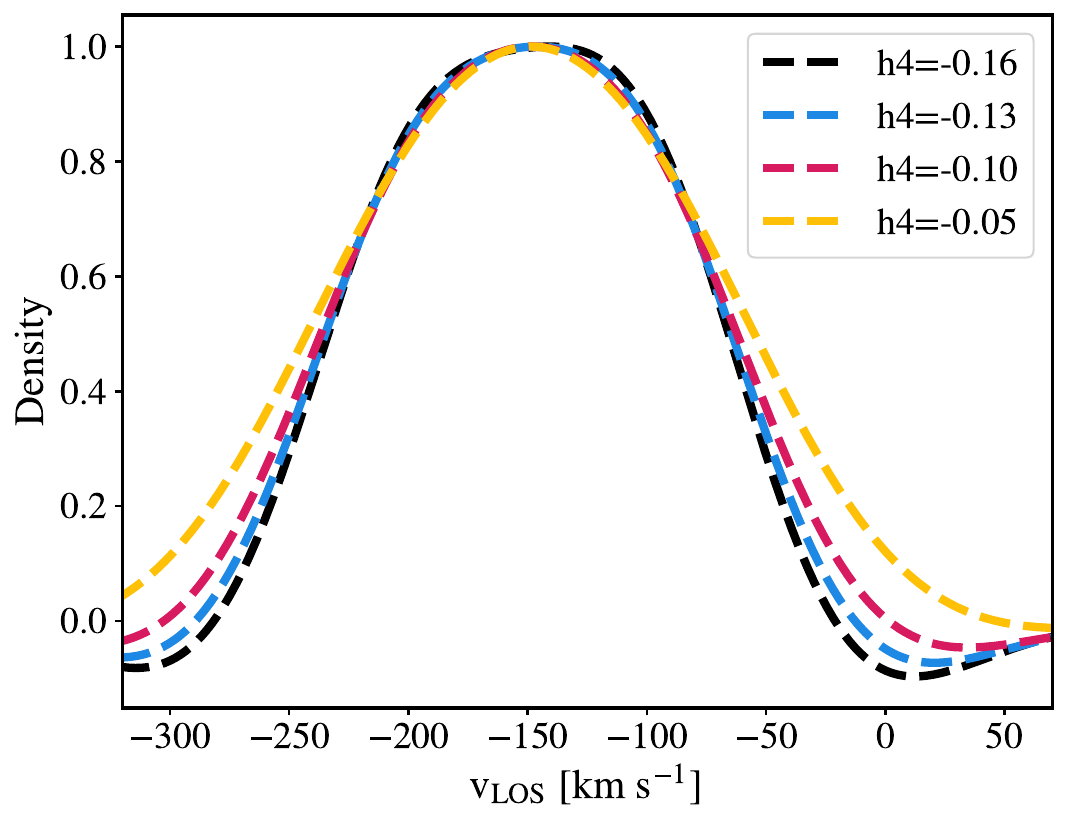}
    
    \end{subfigure}
    
    \caption{Illustration of the effect of velocity resolution on the shape of LOSVDs (left panels) and their Gauss-Hermite fits (right panels), with lines corresponding to LOSVDs smoothed with a Gaussian kernel of 10, 20, 30 and 50 \kms width. The top row corresponds to a region defined by $-4<x<-2.5$ kpc and $-0.5< z<0$ kpc: in the inner disc, the two peaks are far enough apart in velocity to be resolved even at 30 \kms resolution. The bottom row shows the same for the triple-peaked LOSVD from Fig. \ref{fig:galaxy102_combined}: here, the information is lost much faster because peaks are close in velocity. Velocity resolution also affects $h_4$, which shifts closer to zero at low resolution: this effect is much stronger in the case of the three-peaked LOSVD.}
    \label{fig:tripvsdoub}
\end{figure*}

Fig. \ref{fig: resoltion difference} shows how the number of peaks in LOSVDs and the values of $h_4$ change with a reduction in velocity resolution. At a velocity resolution of 20 km s$^{-1}$ most of the structure of the peaks map remains; however, the triple-peaked LOSVDs are no longer present along with some double-peaked regions also disappearing. To explore why this is the case, we investigate how the triple-peaked LOSVD first shown in Fig. \ref{fig:galaxy102_combined} changes with decreasing velocity resolution (see bottom left panel in Fig. \ref{fig:tripvsdoub}). The peaks are separated by only small velocity differences and can no longer be resolved when the velocity resolution is above 10 kms$^{-1}$: the LOSVD becomes double peaked at 20 km s$^{-1}$, and beyond this becomes single-peaked. With the reduction in velocity resolution, the wings also become broader.

Double peaks are usually easier to resolve than triple peaks if they are separate enough in velocity. At a velocity resolution of 20 \kms, double-peaked LOSVDs can be resolved across most of the disc (second row of Fig. \ref{fig: resoltion difference}). By contrast, the velocity resolution is reduced to $30$ kms$^{-1}$, only a small number of double-peaked LOSVDs remain in the central regions of the galaxy, where the velocity difference between peaks is the largest because of projection effects (see an example if the top left panel of Fig. \ref{fig:tripvsdoub}). Finally, at a resolution of $50$ km s$^{-1}$ all LOSVDs become single-peaked.

In contrast, we can see in Fig. \ref{fig: resoltion difference} that the shape of the negative $h_4$ region in the map remains almost consistent as the resolution decreases. The only significant change is that the values of $h_4$ move closer to $0$. To explain why this happens, in the right column of Fig. \ref{fig:tripvsdoub}, we show how the Gauss-Hermite fit to LOSVDs changes with decreasing velocity resolution, for both the double-peaked and triple-peaked LOSVD. As velocity resolution decreases, the wings on Gauss-Hermite fits become broader and the distributions become slightly more peaked. This pushes the values of $h_4$ closer to $0$ and make the signals weaker in general. However, the Gauss-Hermite fits do not vary drastically and all $h_4$ values still remain negative.

Overall, it is evident that the multiple-peaked LOSVDs are harder to detect with decreasing velocity resolution. MUSE will unlikely resolve multiple peaked LOSVDs that are due to spiral arms, but this should be achievable for WEAVE and KOALA at resolutions of $15$ - $20$ km s$^{-1}$. On the other hand, spiral signatures may remain in the $h_4$ maps up to 50 km s$^{-1}$ resolution, and could be detected with MUSE, within the GECKOS survey for example \citep{2024jesse, Fraser_McKelvie_2025}.
If MUSE were to detect regions of negative $h_4$ in the disc of a nearby edge-on galaxy, this could then be followed-up with KOALA or WEAVE in order to resolve multiple peaks in the LOSVDs and confirm the presence of spiral arms.

In practice, detecting the imprint of spiral arms on observed LOSVDs might be further complicated by the presence of dust, given that many galaxies in GECKOS show prominent dust lanes. The effect of dust on the observed kinematics of edge-on galaxies is complex: stars that are closest to us along the line of sight experience little attenuation, while the light from stars that are in the far side of the galaxy might be entirely blocked. Even when assuming a smooth distribution for the dust, this strongly biases the observed kinematics \citep{Bosma1992,Baes2003, Kregel2005}. A patchy distribution of dust, concentrated within spiral arms, will further complicate the interpretation of the observed LOSVDs: the effect of dust is easily seen in maps of $V$, $\sigma$, $h_3$ and $h_4$ shown in \cite{Fraser_McKelvie_2025}, and will also be present when extracting non-parametric LOSVDs for these galaxies. A better understanding of these effects would require using a large set of simulations,  creating mock IFU observations including the effect of dust (as for instance done by \citealp{Barrientos2023}), and running them through a code such as BAYES-LOSVD to see if and when peaks can be detected, which we leave for future work.

\section{Conclusion}
LOSVDs encode important information on the internal structure and formation histories of galaxies. With the advent of GECKOS, a deep MUSE survey of nearby edge-on disc galaxies, and given the availability of new codes to extract non-parametric LOSVDs from observed spectra \citep{2021Falcon, 2026Prashin}, it becomes crucial to better understand the shape of LOSVDs in edge-on discs. In this paper, we use a simulated galaxy to characterise how often we see multiple-peaked LOSVDs in edge-on discs and uncover their origin.

We found that many of the complex LOSVDs exist close to the midplane and are formed as a result of the presence of spiral arms in the disc. The formation of a separate peak is not due to the spiral arms possessing different kinematics from that of the surrounding disc. Instead, it is a result of projection effects combined with the over-densities within the spiral arms. 

We showed how this might be used to quantify the strength of spiral arms above the midplane, by studying how the number of peaks in LOSVDs decreases as a function of height. We also found that the thickness of spiral arms could change as a function of galactocentric radius: in this galaxy, the section of the spiral arm that exists at a large galactocentric radius is thicker than the arms in the inner disc. This leaves clear signatures in the PVDs measured at different heights.

We also explored the effect of spiral arms on the Gauss-Hermite coefficients measured across the galaxy. We compared the multiple-peaked regions with the Gauss-Hermite coefficients and found that $h_4$ is negative in the regions where there are multiple-peaked LOSVDs, hence tracing the spiral arms. We found that the value of $h_4$ is lower when peaks are more numerous and are closer together in velocity.

Finally, we explored how much information is retained in the $h_4$ and peaks map when the velocity resolution of the LOSVDs is  degraded to that of current spectrographs such as MUSE, WEAVE and KOALA. MUSE will unlikely resolve multiple peaked LOSVDs that are due to spiral arms, but this should be achievable for WEAVE and KOALA. On the other hand, spiral signatures in the $h_4$ maps could be detected with MUSE.
If MUSE were to detect regions of negative $h_4$ in the disc of a nearby edge-on galaxy, this could then be followed-up with KOALA or WEAVE in order to resolve multiple peaks in the LOSVDs and confirm the presence of spiral arms.

A full assessment of the observability of spiral arm signatures in LOSVDs of edge-on galaxies would require creating mock IFU observations of a large  set of simulations (including the effect of dust), and running them through a code such as BAYES-LOSVD to see if and when peaks can be detected, which we leave for future work.

\section*{Acknowledgements}
We thank the referee for useful comments. ST acknowledges support from an STFC studentship and from the Bell Burnell Graduate Scholarship Fund. MM acknowledges support from the UK Science and Technology Facilities Council through grant ST/Y002490/1. RJJG acknowledges support from an STFC Ernest Rutherford Fellowship (ST/W003643/1). JF-B acknowledges support from the PID2022-140869NB-I00 grant from the Spanish Ministry of Science and Innovation.

\section*{Data Availability}

The simulation used in this paper is available upon request.



\bibliographystyle{mnras}
\bibliography{Bibliography} 

\begin{thebibliography}{}
\makeatletter
\relax
\def\mn@urlcharsother{\let\do\@makeother \do\$\do\&\do\#\do\^\do\_\do\%\do\~}
\def\mn@doi{\begingroup\mn@urlcharsother \@ifnextchar [ {\mn@doi@} {\mn@doi@[]}}
\def\mn@doi@[#1]#2{\def\@tempa{#1}\ifx\@tempa\@empty \href {http://dx.doi.org/#2} {doi:#2}\else \href {http://dx.doi.org/#2} {#1}\fi \endgroup}
\def\mn@eprint#1#2{\mn@eprint@#1:#2::\@nil}
\def\mn@eprint@arXiv#1{\href {http://arxiv.org/abs/#1} {{\tt arXiv:#1}}}
\def\mn@eprint@dblp#1{\href {http://dblp.uni-trier.de/rec/bibtex/#1.xml} {dblp:#1}}
\def\mn@eprint@#1:#2:#3:#4\@nil{\def\@tempa {#1}\def\@tempb {#2}\def\@tempc {#3}\ifx \@tempc \@empty \let \@tempc \@tempb \let \@tempb \@tempa \fi \ifx \@tempb \@empty \def\@tempb {arXiv}\fi \@ifundefined {mn@eprint@\@tempb}{\@tempb:\@tempc}{\expandafter \expandafter \csname mn@eprint@\@tempb\endcsname \expandafter{\@tempc}}}

\bibitem[\protect\citeauthoryear{{Bacon} et~al.,}{{Bacon} et~al.}{2010}]{2010Bacon}
{Bacon} R.,  et~al., 2010, in {McLean} I.~S.,  {Ramsay} S.~K.,   {Takami} H.,  eds,  Society of Photo-Optical Instrumentation Engineers (SPIE) Conference Series Vol. 7735, Ground-based and Airborne Instrumentation for Astronomy III. p. 773508 (\mn@eprint {arXiv} {2211.16795}), \mn@doi{10.1117/12.856027}

\bibitem[\protect\citeauthoryear{{Baes} et~al.,}{{Baes} et~al.}{2003}]{Baes2003}
{Baes} M.,  et~al., 2003, \mn@doi [\mnras] {10.1046/j.1365-8711.2003.06770.x}, \href {https://ui.adsabs.harvard.edu/abs/2003MNRAS.343.1081B} {343, 1081}

\bibitem[\protect\citeauthoryear{{Barrientos Acevedo} et~al.,}{{Barrientos Acevedo} et~al.}{2023}]{Barrientos2023}
{Barrientos Acevedo} D.,  et~al., 2023, \mn@doi [\mnras] {10.1093/mnras/stad1771}, \href {https://ui.adsabs.harvard.edu/abs/2023MNRAS.524..907B} {524, 907}

\bibitem[\protect\citeauthoryear{{Bosma}, {Byun}, {Freeman}  \& {Athanassoula}}{{Bosma} et~al.}{1992}]{Bosma1992}
{Bosma} A.,  {Byun} Y.,  {Freeman} K.~C.,   {Athanassoula} E.,  1992, \mn@doi [\apjl] {10.1086/186639}, \href {https://ui.adsabs.harvard.edu/abs/1992ApJ...400L..21B} {400, L21}

\bibitem[\protect\citeauthoryear{{Bournaud} \& {Combes}}{{Bournaud} \& {Combes}}{2002}]{ParticleMesh2002}
{Bournaud} F.,  {Combes} F.,  2002, \mn@doi [\aap] {10.1051/0004-6361:20020920}, \href {https://ui.adsabs.harvard.edu/abs/2002A&A...392...83B} {392, 83}

\bibitem[\protect\citeauthoryear{{Bureau} \& {Athanassoula}}{{Bureau} \& {Athanassoula}}{2005}]{Bureau2005}
{Bureau} M.,  {Athanassoula} E.,  2005, \mn@doi [\apj] {10.1086/430056}, \href {https://ui.adsabs.harvard.edu/abs/2005ApJ...626..159B} {626, 159}

\bibitem[\protect\citeauthoryear{Cappellari}{Cappellari}{2025}]{Cappellari2025}
Cappellari M.,  2025, \mn@doi [MNRAS] {10.1093/mnras/staf1726}, 544, 1432

\bibitem[\protect\citeauthoryear{Chung \& Bureau}{Chung \& Bureau}{2004}]{Chung_2004}
Chung A.,  Bureau M.,  2004, \mn@doi [AJ] {10.1086/420988}, 127, 3192

\bibitem[\protect\citeauthoryear{{Davis} et~al.,}{{Davis} et~al.}{2013}]{2013Davis}
{Davis} T.~A.,  et~al., 2013, \mn@doi [\mnras] {10.1093/mnras/sts353}, \href {https://ui.adsabs.harvard.edu/abs/2013MNRAS.429..534D} {429, 534}

\bibitem[\protect\citeauthoryear{{Dobbs} \& {Baba}}{{Dobbs} \& {Baba}}{2014}]{2014Dobbs}
{Dobbs} C.,  {Baba} J.,  2014, \mn@doi [\pasa] {10.1017/pasa.2014.31}, \href {https://ui.adsabs.harvard.edu/abs/2014PASA...31...35D} {31, e035}

\bibitem[\protect\citeauthoryear{{Ellis} et~al.,}{{Ellis} et~al.}{2012}]{2012Ellis}
{Ellis} S.~C.,  et~al., 2012, in {McLean} I.~S.,  {Ramsay} S.~K.,   {Takami} H.,  eds,  Society of Photo-Optical Instrumentation Engineers (SPIE) Conference Series Vol. 8446, Ground-based and Airborne Instrumentation for Astronomy IV. p. 84460V, \mn@doi{10.1117/12.925812}

\bibitem[\protect\citeauthoryear{{Falc{\'o}n-Barroso} \& {Martig}}{{Falc{\'o}n-Barroso} \& {Martig}}{2021}]{2021Falcon}
{Falc{\'o}n-Barroso} J.,  {Martig} M.,  2021, \mn@doi [\aap] {10.1051/0004-6361/202039624}, \href {https://ui.adsabs.harvard.edu/abs/2021A&A...646A..31F} {646, A31}

\bibitem[\protect\citeauthoryear{{Fisher}}{{Fisher}}{1997}]{Fisher}
{Fisher} D.,  1997, \mn@doi [\aj] {10.1086/118312}, \href {https://ui.adsabs.harvard.edu/abs/1997AJ....113..950F} {113, 950}

\bibitem[\protect\citeauthoryear{Fraser-McKelvie et~al.,}{Fraser-McKelvie et~al.}{2025}]{Fraser_McKelvie_2025}
Fraser-McKelvie A.,  et~al., 2025, \mn@doi [A\&A] {10.1051/0004-6361/202452891}, 700, A237

\bibitem[\protect\citeauthoryear{{Fujii}, {Baba}, {Saitoh}, {Makino}, {Kokubo}  \& {Wada}}{{Fujii} et~al.}{2011}]{2011Fujii}
{Fujii} M.~S.,  {Baba} J.,  {Saitoh} T.~R.,  {Makino} J.,  {Kokubo} E.,   {Wada} K.,  2011, \mn@doi [\apj] {10.1088/0004-637X/730/2/109}, \href {https://ui.adsabs.harvard.edu/abs/2011ApJ...730..109F} {730, 109}

\bibitem[\protect\citeauthoryear{{Gebhardt} et~al.,}{{Gebhardt} et~al.}{2000}]{2000Gebhardt}
{Gebhardt} K.,  et~al., 2000, \mn@doi [\aj] {10.1086/301240}, \href {https://ui.adsabs.harvard.edu/abs/2000AJ....119.1157G} {119, 1157}

\bibitem[\protect\citeauthoryear{{Gerhard}}{{Gerhard}}{1993}]{1993Gerhard}
{Gerhard} O.~E.,  1993, \mn@doi [\mnras] {10.1093/mnras/265.1.213}, \href {https://ui.adsabs.harvard.edu/abs/1993MNRAS.265..213G} {265, 213}

\bibitem[\protect\citeauthoryear{{Grand}, {Bovy}, {Kawata}, {Hunt}, {Famaey}, {Siebert}, {Monari}  \& {Cropper}}{{Grand} et~al.}{2015}]{2015Grand}
{Grand} R. J.~J.,  {Bovy} J.,  {Kawata} D.,  {Hunt} J. A.~S.,  {Famaey} B.,  {Siebert} A.,  {Monari} G.,   {Cropper} M.,  2015, \mn@doi [\mnras] {10.1093/mnras/stv1785}, \href {https://ui.adsabs.harvard.edu/abs/2015MNRAS.453.1867G} {453, 1867}

\bibitem[\protect\citeauthoryear{{Gu{\'e}rou}, {Emsellem}, {Krajnovi{\'c}}, {McDermid}, {Contini}  \& {Weilbacher}}{{Gu{\'e}rou} et~al.}{2016}]{Guerou}
{Gu{\'e}rou} A.,  {Emsellem} E.,  {Krajnovi{\'c}} D.,  {McDermid} R.~M.,  {Contini} T.,   {Weilbacher} P.~M.,  2016, \mn@doi [\aap] {10.1051/0004-6361/201628743}, \href {https://ui.adsabs.harvard.edu/abs/2016A&A...591A.143G} {591, A143}

\bibitem[\protect\citeauthoryear{Iannuzzi \& Athanassoula}{Iannuzzi \& Athanassoula}{2015}]{Iannuzzi_2015}
Iannuzzi F.,  Athanassoula E.,  2015, \mn@doi [MNRAS] {10.1093/mnras/stv764}, 450, 2514

\bibitem[\protect\citeauthoryear{{Jethwa}, {Hubmer}, {Ramlau}  \& {Van de Ven}}{{Jethwa} et~al.}{2026}]{2026Prashin}
{Jethwa} P.,  {Hubmer} S.,  {Ramlau} R.,   {Van de Ven} G.,  2026, \mn@doi [\aap] {10.1051/0004-6361/202557989}, \href {https://ui.adsabs.harvard.edu/abs/2026A&A...710A.135J} {710, A135}

\bibitem[\protect\citeauthoryear{{Jim{\'e}nez-L{\'o}pez}, {Garc{\'\i}a-Burillo}, {Querejeta}, {Usero}  \& {Tarr{\'\i}o}}{{Jim{\'e}nez-L{\'o}pez} et~al.}{2026}]{jim}
{Jim{\'e}nez-L{\'o}pez} D.,  {Garc{\'\i}a-Burillo} S.,  {Querejeta} M.,  {Usero} A.,   {Tarr{\'\i}o} P.,  2026, \mn@doi [\aap] {10.1051/0004-6361/202659732}, \href {https://ui.adsabs.harvard.edu/abs/2026A&A...712A..89J} {712, A89}

\bibitem[\protect\citeauthoryear{{Jin} et~al.,}{{Jin} et~al.}{2024}]{2024Jin}
{Jin} S.,  et~al., 2024, \mn@doi [\mnras] {10.1093/mnras/stad557}, \href {https://ui.adsabs.harvard.edu/abs/2024MNRAS.530.2688J} {530, 2688}

\bibitem[\protect\citeauthoryear{Johnston, Merrifield, Arag{\'o}n-Salamanca  \& Cappellari}{Johnston et~al.}{2013}]{johnston2013disentangling}
Johnston E.~J.,  Merrifield M.~R.,  Arag{\'o}n-Salamanca A.,   Cappellari M.,  2013, MNRAS, 428, 1296

\bibitem[\protect\citeauthoryear{{Katkov}, {Sil'chenko}  \& {Afanasiev}}{{Katkov} et~al.}{2013}]{katkov2013}
{Katkov} I.~Y.,  {Sil'chenko} O.~K.,   {Afanasiev} V.~L.,  2013, \mn@doi [\apj] {10.1088/0004-637X/769/2/105}, \href {https://ui.adsabs.harvard.edu/abs/2013ApJ...769..105K} {769, 105}

\bibitem[\protect\citeauthoryear{{Kennicutt}}{{Kennicutt}}{1998}]{kennicutt}
{Kennicutt} Jr. R.~C.,  1998, \mn@doi [\apj] {10.1086/305588}, \href {https://ui.adsabs.harvard.edu/abs/1998ApJ...498..541K} {498, 541}

\bibitem[\protect\citeauthoryear{{Kregel} \& {van der Kruit}}{{Kregel} \& {van der Kruit}}{2004}]{Kregel2004}
{Kregel} M.,  {van der Kruit} P.~C.,  2004, \mn@doi [\mnras] {10.1111/j.1365-2966.2004.07978.x}, \href {https://ui.adsabs.harvard.edu/abs/2004MNRAS.352..787K} {352, 787}

\bibitem[\protect\citeauthoryear{{Kregel} \& {van der Kruit}}{{Kregel} \& {van der Kruit}}{2005}]{Kregel2005}
{Kregel} M.,  {van der Kruit} P.~C.,  2005, \mn@doi [\mnras] {10.1111/j.1365-2966.2005.08811.x}, \href {https://ui.adsabs.harvard.edu/abs/2005MNRAS.358..481K} {358, 481}

\bibitem[\protect\citeauthoryear{{Lamprecht} et~al.,}{{Lamprecht} et~al.}{2026}]{Lamprecht2026}
{Lamprecht} J.,  et~al., 2026, \mn@doi [\aap] {10.1051/0004-6361/202558112}, \href {https://ui.adsabs.harvard.edu/abs/2026A&A...706A.373L} {706, A373}

\bibitem[\protect\citeauthoryear{{Li}, {Shen}, {Bureau}, {Zhou}, {Du}  \& {Debattista}}{{Li} et~al.}{2018}]{Zhao-lu}
{Li} Z.-Y.,  {Shen} J.,  {Bureau} M.,  {Zhou} Y.,  {Du} M.,   {Debattista} V.~P.,  2018, \mn@doi [\apj] {10.3847/1538-4357/aaa771}, \href {https://ui.adsabs.harvard.edu/abs/2018ApJ...854...65L} {854, 65}

\bibitem[\protect\citeauthoryear{{Martig} \& {Bournaud}}{{Martig} \& {Bournaud}}{2010}]{2010Martig}
{Martig} M.,  {Bournaud} F.,  2010, \mn@doi [\apjl] {10.1088/2041-8205/714/2/L275}, \href {https://ui.adsabs.harvard.edu/abs/2010ApJ...714L.275M} {714, L275}

\bibitem[\protect\citeauthoryear{{Martig}, {Bournaud}, {Teyssier}  \& {Dekel}}{{Martig} et~al.}{2009}]{Marie2009}
{Martig} M.,  {Bournaud} F.,  {Teyssier} R.,   {Dekel} A.,  2009, \mn@doi [\apj] {10.1088/0004-637X/707/1/250}, \href {https://ui.adsabs.harvard.edu/abs/2009ApJ...707..250M} {707, 250}

\bibitem[\protect\citeauthoryear{{Martig}, {Bournaud}, {Croton}, {Dekel}  \& {Teyssier}}{{Martig} et~al.}{2012}]{Martig2012}
{Martig} M.,  {Bournaud} F.,  {Croton} D.~J.,  {Dekel} A.,   {Teyssier} R.,  2012, \mn@doi [\apj] {10.1088/0004-637X/756/1/26}, \href {https://ui.adsabs.harvard.edu/abs/2012ApJ...756...26M} {756, 26}

\bibitem[\protect\citeauthoryear{{Martig}, {Minchev}  \& {Flynn}}{{Martig} et~al.}{2014}]{Marie2014}
{Martig} M.,  {Minchev} I.,   {Flynn} C.,  2014, \mn@doi [\mnras] {10.1093/mnras/stu1003}, \href {https://ui.adsabs.harvard.edu/abs/2014MNRAS.442.2474M} {442, 2474}

\bibitem[\protect\citeauthoryear{{Martig} et~al.,}{{Martig} et~al.}{2021}]{Martig2021}
{Martig} M.,  et~al., 2021, \mn@doi [\mnras] {10.1093/mnras/stab2729}, \href {https://ui.adsabs.harvard.edu/abs/2021MNRAS.508.2458M} {508, 2458}

\bibitem[\protect\citeauthoryear{{Mehrgan}, {Thomas}, {Saglia}, {Parikh}  \& {Bender}}{{Mehrgan} et~al.}{2023}]{2023Merghan}
{Mehrgan} K.,  {Thomas} J.,  {Saglia} R.,  {Parikh} T.,   {Bender} R.,  2023, \mn@doi [\apj] {10.3847/1538-4357/acbf2e}, \href {https://ui.adsabs.harvard.edu/abs/2023ApJ...948...79M} {948, 79}

\bibitem[\protect\citeauthoryear{Molaeinezhad, Falcón-Barroso, Martínez-Valpuesta, Khosroshahi, Balcells  \& Peletier}{Molaeinezhad et~al.}{2015}]{Molaeinezhad_2015}
Molaeinezhad A.,  Falcón-Barroso J.,  Martínez-Valpuesta I.,  Khosroshahi H.~G.,  Balcells M.,   Peletier R.~F.,  2015, \mn@doi [MNRAS] {10.1093/mnras/stv2697}, 456, 692

\bibitem[\protect\citeauthoryear{{Monari}, {Famaey}  \& {Siebert}}{{Monari} et~al.}{2016}]{2016Monari}
{Monari} G.,  {Famaey} B.,   {Siebert} A.,  2016, \mn@doi [\mnras] {10.1093/mnras/stw171}, \href {https://ui.adsabs.harvard.edu/abs/2016MNRAS.457.2569M} {457, 2569}

\bibitem[\protect\citeauthoryear{{Peng}, {Ho}, {Impey}  \& {Rix}}{{Peng} et~al.}{2010}]{Peng2010}
{Peng} C.~Y.,  {Ho} L.~C.,  {Impey} C.~D.,   {Rix} H.-W.,  2010, \mn@doi [\aj] {10.1088/0004-6256/139/6/2097}, \href {https://ui.adsabs.harvard.edu/abs/2010AJ....139.2097P} {139, 2097}

\bibitem[\protect\citeauthoryear{{Pinna} et~al.,}{{Pinna} et~al.}{2019}]{Pinna}
{Pinna} F.,  et~al., 2019, \mn@doi [\aap] {10.1051/0004-6361/201833193}, \href {https://ui.adsabs.harvard.edu/abs/2019A&A...623A..19P} {623, A19}

\bibitem[\protect\citeauthoryear{{Rix} \& {White}}{{Rix} \& {White}}{1992}]{1992Rix}
{Rix} H.-W.,  {White} S. D.~M.,  1992, \mn@doi [\mnras] {10.1093/mnras/254.3.389}, \href {https://ui.adsabs.harvard.edu/abs/1992MNRAS.254..389R} {254, 389}

\bibitem[\protect\citeauthoryear{{Rix}, {Franx}, {Fisher}  \& {Illingworth}}{{Rix} et~al.}{1992}]{Rix1992}
{Rix} H.-W.,  {Franx} M.,  {Fisher} D.,   {Illingworth} G.,  1992, \mn@doi [\apjl] {10.1086/186635}, \href {https://ui.adsabs.harvard.edu/abs/1992ApJ...400L...5R} {400, L5}

\bibitem[\protect\citeauthoryear{{Saha} \& {Williams}}{{Saha} \& {Williams}}{1994}]{1994saha}
{Saha} P.,  {Williams} T.~B.,  1994, \mn@doi [\aj] {10.1086/116942}, \href {https://ui.adsabs.harvard.edu/abs/1994AJ....107.1295S} {107, 1295}

\bibitem[\protect\citeauthoryear{{Scott}}{{Scott}}{2015}]{scott}
{Scott} D.~W.,  2015, {Multivariate Density Estimation: Theory, Practice, and Visualization}.
John Wiley and Sons

\bibitem[\protect\citeauthoryear{{Silverman}}{{Silverman}}{1986}]{Silverman}
{Silverman} B.~W.,  1986, {Density estimation for statistics and data analysis}.
Chapman and Hall (London)

\bibitem[\protect\citeauthoryear{{Teyssier}}{{Teyssier}}{2002}]{ADAPTIVE}
{Teyssier} R.,  2002, \mn@doi [\aap] {10.1051/0004-6361:20011817}, \href {https://ui.adsabs.harvard.edu/abs/2002A&A...385..337T} {385, 337}

\bibitem[\protect\citeauthoryear{Virtanen et~al.,}{Virtanen et~al.}{2020}]{Virtanen2020}
Virtanen P.,  et~al., 2020, \mn@doi [Nature Methods] {10.1038/s41592-019-0686-2}, \href {https://rdcu.be/b08Wh} {17, 261}

\bibitem[\protect\citeauthoryear{{Wagner}, {Dettmar}  \& {Bender}}{{Wagner} et~al.}{1989}]{Wagner}
{Wagner} S.~J.,  {Dettmar} R.~J.,   {Bender} R.,  1989, \aap, \href {https://ui.adsabs.harvard.edu/abs/1989A&A...215..243W} {215, 243}

\bibitem[\protect\citeauthoryear{{Wang}, {Sharma}, {Hayden}, {van de Sande}, {Bland-Hawthorn}, {Vaughan}, {Martig}  \& {Pinna}}{{Wang} et~al.}{2024}]{wang2024}
{Wang} Z.,  {Sharma} S.,  {Hayden} M.~R.,  {van de Sande} J.,  {Bland-Hawthorn} J.,  {Vaughan} S.,  {Martig} M.,   {Pinna} F.,  2024, \mn@doi [\mnras] {10.1093/mnras/stae2148}, \href {https://ui.adsabs.harvard.edu/abs/2024MNRAS.534.1175W} {534, 1175}

\bibitem[\protect\citeauthoryear{W{\k{e}}glarczyk}{W{\k{e}}glarczyk}{2018}]{wkeglarczyk2018kernel}
W{\k{e}}glarczyk S.,  2018, in ITM web of conferences. p. 00037

\bibitem[\protect\citeauthoryear{{van de Sande}, {Fraser-McKelvie}, {Fisher}, {Martig}, {Hayden}  \& {Geckos Survey Collaboration}}{{van de Sande} et~al.}{2024}]{2024jesse}
{van de Sande} J.,  {Fraser-McKelvie} A.,  {Fisher} D.~B.,  {Martig} M.,  {Hayden} M.~R.,   {Geckos Survey Collaboration} 2024, in {Tabatabaei} F.,  {Barbuy} B.,   {Ting} Y.-S.,  eds,  IAU Symposium Vol. 377, Early Disk-Galaxy Formation from JWST to the Milky Way. pp 27--33 (\mn@eprint {arXiv} {2306.00059}), \mn@doi{10.1017/S1743921323001138}

\bibitem[\protect\citeauthoryear{{van der Marel} \& {Franx}}{{van der Marel} \& {Franx}}{1993}]{van_der_Marel}
{van der Marel} R.~P.,  {Franx} M.,  1993, \mn@doi [\apj] {10.1086/172534}, \href {https://ui.adsabs.harvard.edu/abs/1993ApJ...407..525V} {407, 525}

\makeatother
\end{thebibliography}




\appendix

\section{Kinematic maps}
\begin{figure}
    \centering
    \begin{subfigure}{\linewidth}
        \centering
        \includegraphics[width=\linewidth]{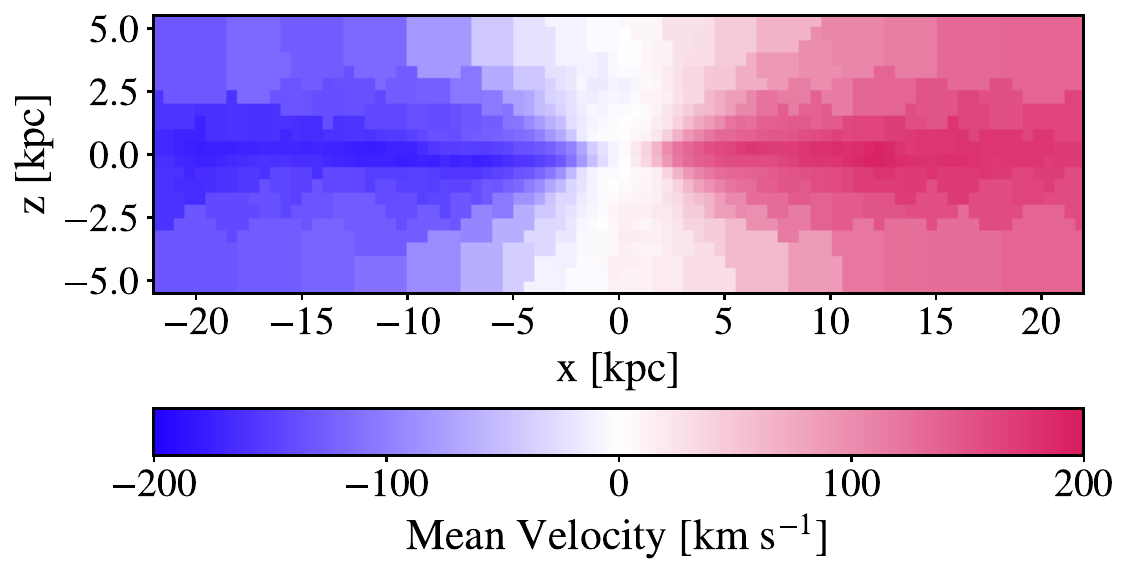}
    \end{subfigure}
    \hfill
    \begin{subfigure}{\linewidth}
        \centering
        \includegraphics[width=\linewidth]{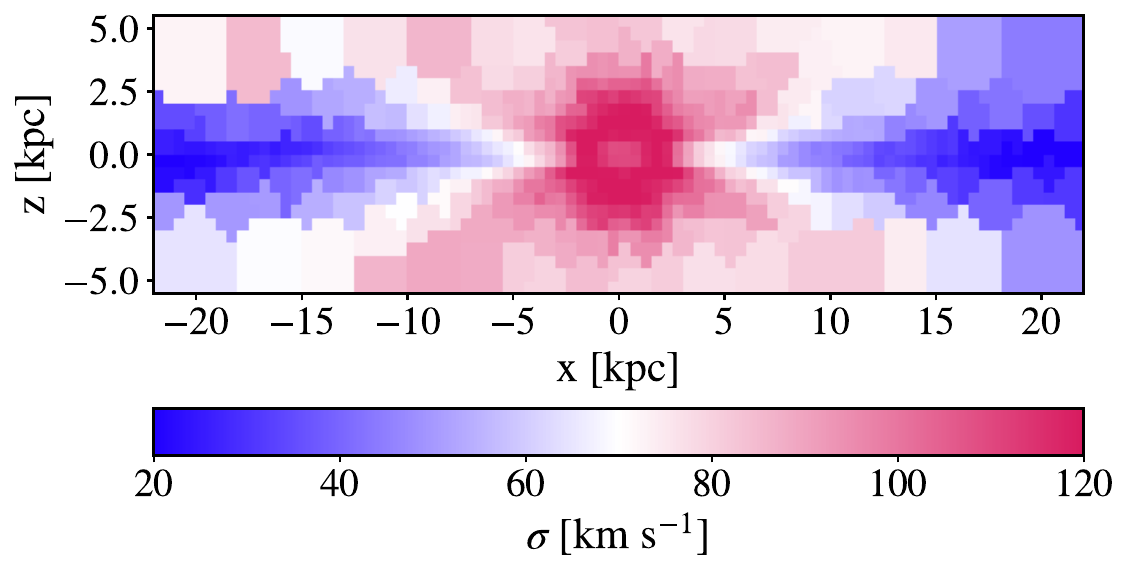}
    \end{subfigure}
    \begin{subfigure}{\linewidth}
        \centering
        \includegraphics[width=\linewidth]{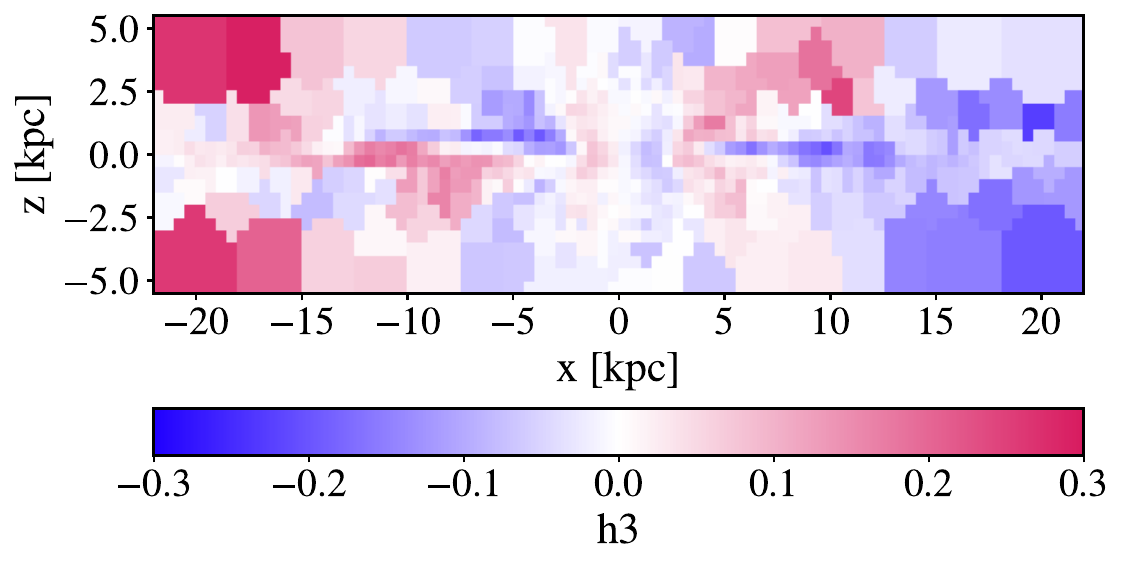}
    \end{subfigure}
    \begin{subfigure}{\linewidth}
        \centering
        \includegraphics[width=\linewidth]{fig3c.pdf}
    \end{subfigure}
        \hfill

    \caption{Kinematic maps of Galaxy 102. The top panel shows the mean velocity map, the second shows the velocity dispersion, the third shows the $h_3$ parameter and the fourth shows the $h_4$ parameter.}
    \label{fig:GHmaps}

\end{figure}

In this Appendix, we present the results from Gauss-Hermite fits to the LOSVDs across the disc of Galaxy 102 (Fig. \ref{fig:GHmaps}).


\bsp	
\label{lastpage}
\end{document}